\documentclass[pra,twocolumn,showpacs]{revtex4}
\usepackage{graphicx}
\usepackage{amsmath}
\usepackage{amssymb}
\usepackage{esint}
\usepackage{color}
\usepackage[normalem]{ulem}

\newcommand{\beq}{\begin{equation}}
\newcommand{\eeq}{\end{equation}}
\newcommand{\beqa}{\begin{eqnarray}}
\newcommand{\eeqa}{\end{eqnarray}}
\newcommand{\ba}{\begin{array}}
\newcommand{\ea}{\end{array}}

\begin{document}

\title{Josephson physics of spin-orbit coupled
elongated Bose-Einstein condensates}
\author{M.A. Garcia-March$^{1}$, G. Mazzarella$^{2}$, L. Dell'Anna$^{2}$,
B. Juli\'a-D\'iaz$^{1,3}$, L. Salasnich$^{2}$, A. Polls$^{1}$}
\affiliation{$^1$Departament d'Estructura i Constituents de la Materia,
Universitat de Barcelona, Diagonal 645, 08028 Barcelona, Spain \\
$^{2}$Dipartimento di Fisica e Astronomia ``Galileo Galilei''
and CNISM, Universit\`a di Padova, Via Marzolo 8, 35131 Padova, Italy\\
$^3$ ICFO–Institut de Ci\`encies Fot\`oniques, Parc Mediterrani de la Tecnologia, 08860 Barcelona, Spain}

\begin{abstract}
We consider an ultracold bosonic binary mixture confined in a
one-dimensional double-well trap.
The two bosonic components are assumed to be two hyperfine internal states of the same atom.
We suppose that these two components are spin-orbit coupled between each other. We employ the two-mode approximation
starting from two coupled Gross-Pitaevskii equations and derive a
system of ordinary differential equations governing the temporal
evolution of the inter-well population imbalance of each component
and that between the two bosonic species. We study the Josephson
oscillations of these spin-orbit coupled Bose-Einstein condensates
by analyzing the interplay between the interatomic interactions
and the spin-orbit coupling and the self-trapped dynamics of the inter-species imbalance. We show that the dynamics of this latter variable is crucially determined by the relationship between the spin-orbit coupling, the tunneling energy, and the interactions.
\end{abstract}

\maketitle


\section{Introduction}

In the last few years artificial spin-orbit (SO) coupling has been
realized in the laboratory with both neutral bosonic systems~\cite{so-bose1}
and fermionic atomic gases~\cite{so-fermi1,so-fermi2}. These achievements
have stimulated theoretical efforts to understand the role of the
SO  with Rashba~\cite{rashba} and Dresselhaus~\cite{dresselhaus}
terms in the physics of the ultracold atoms.

Spin-orbit coupled Bose-Einstein condensates (BECs) have been considered
in Refs.~\cite{stringari1, stringari2}, where the authors have determined
the zero-temperature phase diagram and studied the excitation spectrum
in uniform systems. Bosons in two SO coupled hyperfine states have been
investigated in different contexts, for instance by exploring different
confinements and geometries, e.g. two-dimensional (2D) periodic geometries~\cite{cole},
tight 2D harmonic potential plus a generic one-dimensional (1D)
loose potential~\cite{salasnich1}, and confinements in quasi 1D
parabolic trap~\cite{achilleos}. Also the richer strongly correlated quantum
Hall phases stemming from the spin-orbit coupling (which can be regarded
as a non-abelian external field) have been recently discussed~\cite{tb10,tob13}.
On the fermionic side, the experimental realization of the SO coupling
has produced a growing interest in the study of its role in the crossover from the
Bardeen-Cooper-Schrieffer state of weakly bound Fermi pairs to the BEC of
molecular dimers both for a three-dimensional (3D) uniform Fermi gas~\cite{bcs3d,dms,zhou2} and
in the 2D case~\cite{bcs2d,dms,zhou2}.

One of the richest scenarios opens up when two internal hyperfine states
of the same bosonic atom are coupled between each other
by means of two counter-propagating laser beams and confined by a
one-dimensional double-well. This framework represents the ideal arena
to analyze the atomic counterpart of the Josephson effect which occurs
in the superconductor-oxide-superconductor junctions~\cite{book-barone} with bosonic
binary mixtures~\cite{lobo,xu,satja,diaz1,ajj1,ajj2,rabijosephson,brunonjp}.
It is worth noting that Josephson physics with bosonic mixtures seems within
reach for a number of experimental groups which have in the last years
studied single component Josephson
physics~\cite{Albiez05,GO07,esteve08,gross10,treut10,zib10,amo13,berrada13}.

The aim of the present work is to study the Josephson oscillations with
SO coupled Bose-Einstein condensates. In this case, the
boson-dynamics is ruled by two coupled 3D Gross-Pitaevskii equations, the couplings consisting in the intra- and inter-species interactions
and spin-orbit. By employing the two-spatial mode approximation~\cite{milburn,smerzi},
neglecting the interatomic interactions between bosons in different wells,
and assuming that the Rashba and the Dresselhaus velocities are equal,
one derives a system of ordinary differential equations (ODEs). The
solution of these ODEs provides the temporal evolution of the relevant
variables of the problem. These are the two fractional population imbalances,
$z_k$, defined as the difference in the occupation of the two wells for
each bosonic component labeled by $k=1,2$. On top of these we define
the total population imbalance, $z_{12}$,  between the bosons in species
$1$ and those in species $2$. We term this latter variable the
{\it  polarization}, i.e. zero polarization implies an equal amount of atoms
populating the components $1$ and $2$, while the system is fully polarized
if all atoms populate either of the two species. The dynamical evolution of $z_{12}$
gives the interchange of atoms between the two bosonic components
and is therefore directly related to the SO coupling.

The issue of the Josephson physics in the presence of spin-orbit
has been recently addressed by Zhang and co-workers \cite{zhang}
that (as we comment in Sec. ~\ref{s2}) have analyzed the problem starting
from a quantum single-particle (SP) Hamiltonian different from that we consider in this paper. The Hamiltonian that will used here - to make stronger the link with experiments on SO coupled BECs - is the same considered by Lin {\it et al.} \cite{so-bose1}.

We numerically solve the aforementioned ODEs and study the temporal behavior of the three $z$'s
and their corresponding canonically conjugated phases. We focus on the
interplay between the boson-boson interaction and the inter-well spin-orbit coupling
(the intra-well SO is zero in first approximation) in determining the
importance of the interchange of atoms between the two bosonic species
on the two canonical effects: self-trapping and macroscopic tunneling
phenomena. In particular, we point out that an analogous of the macroscopic
quantum self-trapping that one observes for the variable
$z_{k}$~\cite{satja,ajj2} exists for the polarization $z_{12}$ as well.
During the $z_{12}$ self-trapped dynamics, on the average, one bosonic component is more populated than the other one.

The article is organized in the following way. First, in Sec.~\ref{s2} we
describe the single particle Hamiltonian of the system, following the
experimental realization of Ref.~\cite{so-bose1}. In Sec.~\ref{s3} we
present the mean-field description of the problem. In Sec.~\ref{s4}
two-mode approximation is exploited to derive the ordinary differential equations that we use to describe the dynamics of our system.
In Sec.~\ref{s5} we study the effect of the spin-orbit term on
the macroscopic quantum tunneling and self-trapping. A summary and
conclusions are provided in Sec.~\ref{s6}.

\section{Artificial spin-orbit coupling}
\label{s2}

We consider two counter-propagating laser beams which couple two
internal hyperfine states of the same bosonic atom (e.g., the
$m_F=0,1$ components of an $^{87}$Rb $F=1$ spinor condensate as in the experiment carried out by  Lin and co-workers \cite {so-bose1}) by a resonant
stimulated two-photon Raman transition characterized by a Rabi
frequency $\Omega_R$. This transition can be induced by a laser beam
with a detuning $\delta$ with respect to the spacing between the
energy levels of  the two hyperfine states. The two internal
hyperfine states define the pseudo-spin space of each atom.
The single-particle (SP) quantum Hamiltonian can be written as
\beqa
\label{eq:SP_part}
H_{sp} &=&
\left[ {{\bf p}^2\over 2m}
+ U({\bf r}) \right] \sigma_0 +
v_R \left( p_x \sigma_y - p_y \sigma_x \right)
\nonumber
\\
&+& v_D \left(  p_x \sigma_y + p_y \sigma_x \right) +
{\hbar \Omega_R\over 2} \sigma_z+{\hbar \delta\over 2} \sigma_y
\;,\eeqa
where ${\bf p}=(p_x , p_y , p_z )=
-i\hbar (\partial_x,\partial_y,\partial_z)$ is the linear momentum
operator, $U({\bf r})$ is the external trapping potential,
$v_R$ and $v_D$ are, respectively, the Rashba and Dresselhaus velocities,
$\sigma_0$ is the $2\times 2$ identity matrix, and $\sigma_x$,
$\sigma_y$, $\sigma_z$ are the Pauli matrices. In the recent
experiments~\cite{so-bose1,so-fermi1,so-fermi2} one has
$v_R=v_D \equiv v$, which is the configuration that we shall consider
in the present paper. In this configuration, the terms proportional to
the Pauli matrices in the single-particle Hamiltonian (\ref{eq:SP_part})
are the same appearing in Eq.~(2) of \cite{zhang}, provided to perform on
the SP Hamiltonian of Zhang and co-workers the following global pseudo-spin rotation:
$\sigma_z \rightarrow \sigma_y$, $\sigma_y \rightarrow \sigma_x$, and $\sigma_x \rightarrow \sigma_z$.

\section{Mean-field description: Gross-Pitaevskii equations}
\label{s3}

We consider a dilute binary condensate with a spin-orbit coupling term
as described above, whose single-particle Hamiltonian is given
by Eq.~(\ref{eq:SP_part}). The ultracold atomic cloud is further
assumed to be confined in the transverse $(y,z)$ plane by a strong harmonic
potential with frequency $\omega_{\bot}$ and in the axial ($x$) direction
by a generic weak potential $V(x)$, namely
\beq
U(\mathbf{r}) = V(x) + {1\over 2} m\omega_{\bot}^2 (y^2+z^2) \; .
\label{extrapping}
\eeq
Under the hypothesis that the boson-boson interactions (intra- and inter-species) can be
described by a contact potential, the two time-dependent 3D Gross-Pitaevskii equations (GPEs) which describe the binary condensate are
\[
i\,\partial _{t}\psi_{1}=\Big[ -{\frac{1}{2}}\nabla ^{2}
+ V(x) + {\frac{1}{2}} (y^{2}+z^{2}) + \Gamma + 2\pi \,\tilde g_{1}|\psi_{1}|^{2}
\]%
\begin{equation}
+2\pi \,\tilde g_{12}|\psi _{2}|^{2}\Big]\psi_{1}
- (\gamma \partial_x+i\tilde \delta) \psi_2 \; ,
\label{GPE1}
\end{equation}
\[
i\,\partial _{t}\psi_{2}=\Big[ -{\frac{1}{2}}\nabla ^{2}
+ V(x) + {\frac{1}{2}} (y^{2}+z^{2}) - \Gamma
+ 2\pi\,\tilde g_{2}|\psi_{2}|^{2}
\]
\begin{equation}
+2\pi\,\tilde g_{12}|\psi_{1}|^{2}\Big]
\psi_{2} + (\gamma \partial_x+i\tilde \delta) \psi_{1} \; ,
\label{GPE}
\end{equation}
where lengths, times, and energies are written in units of
$a_{\bot}=\sqrt{\hbar/(m\omega_{\bot })}$, $\omega_{\bot }^{-1}$,
and $\hbar \omega_{\bot }$, respectively. Here $\psi_{k}(x,y,z,t)$
is the macroscopic wave function of the $k$th atomic hyperfine
state ($k=1,2$). The number of atoms populating each hyperfine
state can be computed at any time as
\begin{equation}
\label{norm}
\int \int \int \left\vert \psi _{k}(x,y,z,t)\right\vert ^{2}dx\ dy\ dz
= N_{k}(t) \; .
\end{equation}
The strengths of the intra- and inter-species interactions are
given by
\begin{equation}
\tilde g_{k}\equiv 2a_{k}/a_{\bot },~\tilde g_{12}\equiv 2a_{12}/a_{\bot },  \label{ggg}
\end{equation}%
where $a_k$ and $a_{12}$ are the $s$-wave scattering lengths pertaining, respectively, to the intra-species interaction and inter-species one.

Note that $\gamma = 2v/(a_{\bot}\omega_{\bot})$ is the dimensionless
SO coupling, $\Gamma=\Omega_R/(2\omega_{\bot})$ and
$\tilde \delta=\delta/(2\omega_{\bot})$ are the dimensionless Rabi
and detuning frequencies, respectively.

These GPEs can be derived with the ordinary variational procedure
from the Lagrangian density
\beq
L = L_0 + L_{SO} + L_I \; ,
\label{L0}
\eeq
where
\beq
\label{lo}
L_0 =\sum_{k=1,2}\psi _{k}^{\ast }\left[ i\,\partial_{t}
+{\frac{1}{2}}\nabla^{2}-V(x)-{\frac{1}{2}}(y^{2}+z^{2}) \right]\psi_{k} \;
\eeq
is the Lagrangian density of the non-interacting binary condensate,
\beqa
\label{lso}
L_{SO} &=& - \Gamma \left( |\psi_1|^2 - |\psi_2|^2 \right)
+ \gamma \Big( \psi_1^*\partial_x\psi_2
- \psi_2^*\partial_x\psi_1 \Big) \nonumber\\
&+&i\tilde \delta\Big( \psi_1^*\psi_2 - \psi_2^*\psi_1 \Big)
\;\eeqa
is the SO coupling contribution, and
\beq
\label{li}
L_I = - {\pi \,\tilde g_1} |\psi_{1}|^4  - {\pi \,\tilde g_2} |\psi_{2}|^4
-2\pi\,\tilde g_{12}|\psi_{1}|^{2}|\psi_{2}|^{2} \; ,
\eeq
is the term due to $s$-wave interactions between the bosonic atoms.

By assuming a strongly localized cloud in the transverse plane and weakly localized in the axial direction (see the above discussion about the trapping potential (\ref{extrapping})), we can derive a system of effective one-dimensional Gross-Pitaevskii equations by adopting the usual Gaussian ansatz for the wave functions $\psi_k$ ($k=1,2$) in the directions $y$ and $z$, that is
\beq
\psi_{k}(x,y,z,t)={\frac{1}{\sqrt{\pi }a_{\bot}}}
\exp {\left\{ -{\frac{y^{2}+z^2}{2a_{\bot}^{2}}}\right\} }\,f_{k}(x,t)
\; ,
\label{ansatz}
\eeq
where the complex functions $f_{k}(x,t)$ ($k=1,2$) are dynamical fields,
obeying normalization $\int_{-\infty }^{+\infty }\left\vert
f_{k}(x,t)\right\vert ^{2}dx=N_{k}$, as it follows from
Eqs.~(\ref{norm}) and~(\ref{ansatz}).

By inserting the ansatz (\ref{ansatz}) in the Lagrangian density $L$ given by Eq.~(\ref{L0}) (its contributions being the right-hand sides of Eqs.~(\ref{lo})-(\ref{li})) and performing the integration in the transverse directions ($y$ and $z$) starting from $\displaystyle{\int dxdydz L}$ we obtain the effective 1D Lagrangian
\begin{eqnarray}
\label{effl}
\bar L &=& \int\, dx \bigg(\bigg[\sum_{k=1,2} f_{k}^{*}(i\partial_t+\frac{1}{2}\partial_x^{2})f_k\nonumber\\
&-&(1+V(x))|f_k|^2-\frac{\tilde g_k}{2}|f_k|^4\bigg]-\tilde g_{12}\,|f_1|^2|f_2|^2\nonumber\\
&-&\Gamma(|f_1|^2 - |f_2|^2)+\gamma\,(f_1^{*}\partial_xf_2-f_2^{*}\partial_x f_1)\nonumber\\
&+&i\tilde \delta (f_1^*f_2 - f_2^*f_1)\bigg)\;.
\end{eqnarray}
By varying $\bar L$ with respect to $f_{k}^{*}$ we get the following system formed by two coupled 1D GPEs
\beqa
\label{f1eq}
i\,\partial_{t}f_{1} &=& \Big[ -{\frac{1}{2}}\partial_x^{2} + V(x) + \Gamma
+ \tilde g_{1}|f_{1}|^{2}
+ \tilde g_{12}|f_{2}|^{2}\Big] f_{1} \nonumber\\
&-& (\gamma \partial_x+i \tilde \delta )\, f_{2} \; ,
\eeqa
\beqa
\label{f2eq}
i\,\partial_{t}f_{2} &=& \Big[ -{\frac{1}{2}}\partial_x^{2} + V(x) - \Gamma
+ \tilde g_{2}|f_{2}|^{2}+\tilde g_{12}|f_{1}|^{2}\Big] f_{2} \nonumber\\
&+& (\gamma \partial_x+i\tilde \delta) \, f_{1} \; .
\eeqa

\section{Bimodal approximation}
\label{s4}

Let us suppose that the potential $V(x)$ at the right-hand side of Eq.~(\ref{extrapping}) is a symmetric
double-well potential $V_{\mathrm{DW}}(x)$. To describe the dynamics by a finite-mode approximation, we use a two-mode ansatz for each wave function $f_{k}$ (recall $k=1,2$), as
originally introduced in~\cite{milburn}:
\beq
\label{twomodef}
f_{k}(x,t)=\Psi _{k}^{L}(t)\phi _{k}^{L}(x)
+\Psi _{k}^{R}(t)\phi_{k}^{R}(x)\;.
\eeq
The functions $\phi_{k}^{\alpha }(x)$ ($\alpha =L,R$) - that we consider as real functions as done for example in \cite{ajj1,ajj2,rabijosephson} -
are single particle wave functions tightly localized in the $\alpha$th
well, while the time dependence is encoded in
$
\Psi_{k}^{\alpha }(t)\equiv \sqrt{N_{k}^{\alpha }(t)} \
e^{i\theta _{k}^{\alpha}(t)}\;,
$
where the total number of particles in the $k$th species is given by
$N_{k}^{L}(t)+N_{k}^{R}(t)=\left\vert \Psi _{k}^{L}(t)\right\vert ^{2}
+\left\vert\Psi _{k}^{R}(t)\right\vert ^{2}\equiv N_{k}(t)$. The functions $\phi_{k}^{\alpha }(x)$ satisfy the following orthonormalization conditions: $\int dx (\phi _{k}^{\alpha}(x))^2=1$ and $\int dx \phi _{k}^{L}(x)\phi _{k}^{R}(x)=0$.

At this point, we exploit the two-mode approximation for $f_{k}(x,t)$ given by Eq. (\ref{twomodef})
in each of the two coupled one-dimensional Gross-Pitaevskii equations, (\ref{f1eq}) and (\ref{f2eq}). By left-multiplying both the left-hand side and the right-hand one of the $f_k$-1D GPE for $\phi _{k}^{\alpha}(x)$ and integrating in the $x$ direction, one obtains the equations of motion for $N_{k}^{\alpha}$ and $\theta_{k}^{\alpha}$ (note that here one has to take into account the above orthonormalization conditions). We retain only the integrals with
wave functions localized in the same well for the boson-boson interactions and terms related to
the Rabi and detuning terms. Conversely, due to the derivative coupling, both the SO coupling in the same and  different wells have to be considered. The equations of
motion for $N_{k}^{\alpha}$ and $\theta_{k}^{\alpha}$ thus read,
\begin{eqnarray}
\label{eq:Nkalpha}
\dot{N}_{k}^\alpha & = &
- 2J_{k} \sqrt{N_{k}^{\alpha'} N_{k}^{\alpha}}
\sin\left(\theta_{k}^{\alpha'}-\theta_{k}^{\alpha}\right)\nonumber\\
&+& 2 S_{12}^{\alpha}\sqrt{N_{3-k}^{\alpha} N_{k}^{\alpha}}
\sin\left(\theta_{k}^\alpha-\theta_{3-k}^{\alpha}\right)\nonumber\\
&\pm& 2 \bar S_{12}^{\alpha,\alpha'}\sqrt{N_{3-k}^{\alpha'} N_{k}^{\alpha}}
\sin\left(\theta_{k}^\alpha-\theta_{3-k}^{\alpha'}\right)\nonumber\\
&\mp& 2 D_{12}^{\alpha}\sqrt{N_{3-k}^\alpha N_{k}^{\alpha}}
\cos\left(\theta_{k}^\alpha-\theta_{3-k}^\alpha\right)
\;,\end{eqnarray}
\begin{eqnarray}
\label{eq:thekalpha}
\dot{\theta}_{k}^\alpha & = &-\left(U_{k} N_{k}^\alpha
+\epsilon_{k}+U_{12} N_{3-k}^\alpha \pm \Gamma \right) \nonumber\\
&+& J_{k}\sqrt{\frac{N_{k}^{\alpha'}}{N_{k}^\alpha}}
\cos\left(\theta_{k}^\alpha-\theta_{k}^{\alpha'}\right)\nonumber\\
&+& S_{12}^{\alpha}\sqrt{\frac{N_{3-k}^{\alpha}}{N_{k}^\alpha}}
\cos\left(\theta_{k}^\alpha-\theta_{3-k}^{\alpha}\right)\nonumber\\
&\pm&\bar S_{12}^{\alpha,\alpha'}\sqrt{\frac{N_{3-k}^{\alpha'}}{N_{k}^\alpha}}
\cos\left(\theta_{k}^\alpha-\theta_{3-k}^{\alpha'}\right)\nonumber\\
&\pm& D_{12}^{\alpha}\sqrt{\frac{N_{3-k}^\alpha}{N_{k}^\alpha}}
\sin\left(\theta_{k}^\alpha-\theta_{3-k}^\alpha\right).
\end{eqnarray}
%
The sign in front of the $D_{12}$ term in Eq.~(\ref{eq:Nkalpha})
is minus (plus) for $k=1 (2)$. In Eq.~(\ref{eq:thekalpha}) the
sign in front of $\Gamma$ is plus (minus) for $k=1 (2)$. In both equations
the sign $+$ in front of $\bar S_{12}^{\alpha,\alpha'}$ is plus (minus) for the component 1(2).

The following constants are introduced in terms of the
single-particle functions $\phi_k^{\alpha}$:
\begin{eqnarray}
\label{parameters}
&&\epsilon_{k}^{\alpha}=\int dx\,\bigg[\phi_{k}^{\alpha}(x)\,
\big(-\frac{1}{2}\partial^{2}_{x}
+V_{\mathrm{DW}}(x)\big)\,\phi_{k}^{\alpha}(x)\bigg]+1,\nonumber\\
&&U_{k}=\tilde g_k\int dx\,(\phi_{k}^{\alpha})^{4}, \nonumber\\
&&J_{k}=-\int dx\,\bigg[\phi_{k}^{L}(x)\,
\big(-\frac{1}{2}\partial^{2}_{x} +V_{\mathrm{DW}}(x)\big)\,\phi_{k}^{R}(x)\bigg],
\nonumber\\
&&U_{12}=\tilde g_{12}\int dx\,(\phi_{1}^{\alpha})^{2}(\phi_{2}^{\alpha})^{2},\nonumber\\
&& S_{1 2}^{\alpha}=\gamma\,
\int dx\,\phi_{1}^{\alpha}(x)\partial_x \phi_{2}^{\alpha}(x)\;,\nonumber \\
&&\bar S_{1 2}^{\alpha,\alpha'}=\gamma\,
\int dx\,\phi_{1}^{\alpha}(x)\partial_x \phi_{2}^{\alpha'}(x)\;,\nonumber \\
&&D_{12}^{\alpha}=\tilde \delta\int dx\,\phi_{1}^{\alpha}(x)\phi_{2}^{\alpha}(x).
\end{eqnarray}
In obtaining the equations of motion~(\ref{eq:Nkalpha})
and (\ref{eq:thekalpha}), we have used the double-well left-right symmetry
so that: $\epsilon_{k}^{L}=\epsilon_{k}^{R} \equiv \epsilon_k$,
$U_{k}^{L}=U_{k}^{R} \equiv U_k$ and $U_{12}^{L}=U_{12}^{R} \equiv U_{12}$.
 We have employed, moreover, the following properties: by integrating by parts we have both $S_{12}^{\alpha}=- S_{21}^{\alpha}$ and
$\bar S_{12}^{\alpha,\alpha'}=-\bar S_{21}^{\alpha',\alpha}$, and finally that $\bar S_{12}^{\alpha,\alpha'}=-\bar S_{12}^{\alpha',\alpha}$.

We are assuming that $m_1=m_2=m$, thus it is possible to work
with the same localized modes for both components, that is
$\phi_1^{\alpha}=\phi_2^{\alpha}=\phi^{\alpha}$ ($\alpha=L,R$). If this
is the case, $\epsilon_1=\epsilon_2=\epsilon$,  $J_1=J_2=J$, and
$D_{12}=\tilde \delta$. This also implies that $S_{12}^{\alpha}$ and $S_{21}^{\alpha}$ vanish. After  defining $S_+=\bar S_{12}^{L,R}=-\bar S_{21}^{R,L}$ and   $S_-=\bar S_{12}^{R,L}=-\bar S_{21}^{L,R}$,  we can rewrite the  equations of motion~(\ref{eq:Nkalpha})
and (\ref{eq:thekalpha})  as:

\begin{eqnarray}
\label{eq:Nkalpha1}
\dot{N}_{k}^\alpha & = &
- 2J \sqrt{N_{k}^{\alpha'} N_{k}^{\alpha}}
\sin\left(\theta_{k}^{\alpha'}-\theta_{k}^{\alpha}\right)\nonumber\\
&+& 2  S_{\pm(\mp)}\sqrt{N_{3-k}^{R(L)} N_{k}^{L(R)}}
\sin\left(\theta_{k}^{L(R)}-\theta_{3-k}^{R(L)}\right)\nonumber\\
&\mp& 2 \tilde\delta\sqrt{N_{3-k}^\alpha N_{k}^{\alpha}}
\cos\left(\theta_{k}^\alpha-\theta_{3-k}^\alpha\right)
\;,\end{eqnarray}
\begin{eqnarray}
\label{eq:thekalpha1}
\dot{\theta}_{k}^\alpha & = &-\left(U_{k} N_{k}^\alpha
+U_{12} N_{3-k}^\alpha \pm \Gamma \right) \nonumber\\
&+& J\sqrt{\frac{N_{k}^{\alpha'}}{N_{k}^\alpha}}
\cos\left(\theta_{k}^\alpha-\theta_{k}^{\alpha'}\right)\nonumber\\
&+&S_{\pm(\mp)}\sqrt{\frac{N_{3-k}^{R(L)}}{N_{k}^{L(R)}}}\cos\left(\theta_{k}^{L(R)}-\theta_{3-k}^{R(L)}\right)\nonumber\\
&\pm& \tilde\delta\sqrt{\frac{N_{3-k}^\alpha}{N_{k}^\alpha}}
\sin\left(\theta_{k}^\alpha-\theta_{3-k}^\alpha\right).
\end{eqnarray}
The upper (lower) sign in these equations  holds for $k=1(2)$. For the  SO coupling term ($S_\pm$), the symbol without (with) parenthesis  holds  for the equation of motion of the occupation $N_{k}^\alpha$ or $\theta_{k}^\alpha$ with $\alpha=L(R)$. In Appendix~\ref{sec:appendix1} we show how to derive these
equations as the semiclassical limit of the exact bimodal
many-body Hamiltonian.
Note that the equations (\ref{eq:Nkalpha1}) and (\ref{eq:thekalpha1}) can be regarded as obtained from a classical
Hamiltonian, hence being $\theta_{k}^\alpha$ and $N_{k}^\alpha$ the canonical conjugate
variables. Then, $\dot{N}_{k}^\alpha=\partial H/\partial\theta_{k}^\alpha$
and $\dot{\theta}_{k}^\alpha=-\partial H/\partial N_{k}^\alpha$ with a
classical Hamiltonian:
\begin{eqnarray}
\label{semiclassicalH}
H & = & \sum_{k,\alpha}\left\{ \left[\frac{1}{2} U_k N_{k}^\alpha
\pm\Gamma+\frac{1}{2}U_{12} N_{3-k}^\alpha \right] N_{k}^\alpha \right.\nonumber\\
& - & \left. J\sqrt{N_{k}^\alpha N_{k}^{\alpha'}}
\cos\left(\theta_{k}^\alpha-\theta_{k}^{\alpha'}\right)\right\}\nonumber \\
 & - & 2  S_{+}\sqrt{N_{2}^{R} N_{1}^L}
\cos\left(\theta_{1}^L - \theta_{2}^{R} \right)\nonumber \\
 & - & 2 S_{-}\sqrt{N_{2}^{L} N_{1}^R}
\cos\left(\theta_{1}^R - \theta_{2}^{L} \right)
 \nonumber \\
 & - & 2 \sum_{\alpha}\left\{ \tilde \delta \sqrt{N_{1}^\alpha N_{2}^\alpha}
\sin\left(\theta_{1}^\alpha-\theta_{2}^\alpha\right)\right\},
\end{eqnarray}
with  the upper (lower) sign for $k=1 (2)$.
Noting that since there is one constant of motion, that is the total number
of atoms, we can reduce the number of variables through the
following transformation:
\begin{equation}
\label{eq:deftransform}
\left(\begin{array}{c}
1\\
z_{1}\\
z_{2}\\
z_{12}
\end{array}\right)\!=\!\frac{\mathrm{M}}{N}\left(\begin{array}{c}
N_{1}^L\\
N_{1}^R\\
N_{2}^L\\
N_{2}^R
\end{array}\right)\!,\,\,\left(\begin{array}{c}
\theta_{N}\\
\theta_{1}\\
\theta_{2}\\
\theta_{12}
\end{array}\right)\!=\!-\mathrm{M}\left(\begin{array}{c}
\theta_{1}^L\\
\theta_{1}^R\\
\theta_{2}^L\\
\theta_{2}^R
\end{array}\right),
\end{equation}
with
\[
\mathrm{M}=\left(\begin{array}{cccc}
1 & 1 & 1 & 1\\
1 & -1 & 0 & 0\\
0 & 0 & 1 & -1\\
1 & 1 & -1 & -1
\end{array}\right).
\]
The set of new variables $\{z_{i},\theta_{i}\}$ are also canonically
conjugate because their Poisson brackets fulfill
\begin{equation}
\{z_{i},\theta_{j}\}\equiv\sum_{k,\alpha}\left(\frac{\partial z_{i}}{\partial N_{k}^\alpha}\frac{\partial\theta_{j}}{\partial\theta_{k}^\alpha}-\frac{\partial z_{i}}{\partial\theta_{k}^\alpha}\frac{\partial\theta_{j}}{\partial N_{k}^\alpha}\right)=\delta_{ij}.
\end{equation}
From the transformations (\ref{eq:deftransform}), one realizes that
\begin{eqnarray}
\label{zs}
&&z_1=\frac{N_1^{L}-N_1^{R}}{N} \;,
z_2=\frac{N_2^{L}-N_2^{R}}{N} \nonumber\\
&&z_{12}=\frac{N_1-N_2}{N}
\;,\end{eqnarray}
with $N_k=N_k^{L}+N_k^{R}$. The phases associated to the $z_1$, $z_2$, and $z_{12}$ are, respectively
\begin{eqnarray}
\label{thetas}
&& \theta_1=\theta_1^{R}-\theta_1^{L} \;,
 \theta_2=\theta_2^{R}-\theta_2^{L} \nonumber\\
&& \theta_{12}=\theta_2^{L}+\theta_2^{R}-(\theta_1^{L}+\theta_1^{R})
\;.\end{eqnarray}
In the rotated frame \cite{zhang}, $N_{\uparrow}$ and $N_{\downarrow}$, the numbers of bosons in the dressed hyperfine states $\uparrow$ and $\downarrow$, can be written in terms of our quantities as follows
\beq
\label{nupdown}
N_{\sigma}=\frac{1}{2}\big(N_1+N_2\pm2\,\sum_{\alpha=L,R}\sqrt{N_{1}^{\alpha}N_2^{\alpha}}
\cos(\theta_{1}^{\alpha}-\theta_{2}^{\alpha})\big)
\eeq
with plus (minus) for $\sigma=\uparrow(\downarrow)$, and $N_{k}^{\alpha}$ and $\theta_k^{\alpha}$ solutions of Eqs. (\ref{eq:Nkalpha1})-(\ref{eq:thekalpha1}). Accordingly, the population imbalance $z_{\uparrow,\downarrow}=(N_{\uparrow}-N_{\downarrow})/N$ is related to our variables in the following way:
\beq
\label{zupdown}
z_{\uparrow,\downarrow}= \frac{2\big(\sum_{\alpha=L,R}\sqrt{N_{1}^{\alpha}N_2^{\alpha}} \cos(\theta_{1}^{\alpha}-\theta_{2}^{\alpha})\big)}{N}
\;.\eeq 
Remarkably, the variables (\ref{zs}) and (\ref{thetas}) are directly related to the usual
Josephson physics. Namely, $z_k$ is the population imbalance
of component $k$, $\theta_k$ is its corresponding canonical phase. The polarization
$z_{12}$ measures the total population transfer between both components, that is,
the population imbalance between the first and the second species.
In the absence of spin-orbit coupling the variable $z_{12}$ becomes a constant
of motion. Its evolution will thus be intimately related to the
effect of the SO term. $\theta_{12}$ is the canonical phase associated to $z_{12}$.

In terms of these new variables the Hamiltonian governing the dynamics is
$H'=2H/N$ which reads
%
\begin{eqnarray}
\label{semiclassicalH1}
H' \!\!&\! = \!& \!2\Gamma z_{12}-\!2J\sum_k \sqrt{( z_{12}\!\pm\!1)^{2}\!-\!4 z_{k}^{2}}\cos\left(\theta_{k}\right)\\
&\hspace{-0.95cm}\!+\!\!& \hspace{-0.4cm}\!\!\frac{1}{8} N \sum_k U_k \left[( z_{12}\!\pm\!1)^{2}\!+\! 4 z_{k}^{2}\right]\!+\! \frac{1}{4}U_{12} N(1\!+\!4z_{1} z_{2}\!-\!z_{12}^{2}) \nonumber\\
 &\hspace{-0.95cm}\!+\!\!& \hspace{-0.4cm}\!\! 2S_+\! \left[\left(1\!-\!2z_{2}\!-\!z_{12}\right)\!\left(1\!+\!2z_{1}\!+\!z_{12}\right)\right]^{\frac{1}{2}}\!\cos[\frac{1}{2}(\theta_{1}\!+\!\theta_{2}\!+\!\theta_{12})]\nonumber\\
 &\hspace{-0.95cm}\!+\!\!& \hspace{-0.4cm}\!\!  2S_-\! \left[\left(1\!+\!2z_{2}\!-\!z_{12}\right)\!\left(1\!-\!2z_{1}\!+\!z_{12}\right)\right]^{\frac{1}{2}}\!\cos[\frac{1}{2}(\theta_{1}\!+\!\theta_{2}\!-\!\theta_{12})]\nonumber\\
 &\hspace{-0.95cm}\!\mp\!\!& \hspace{-0.4cm}\!\! 2\bar \delta\!\sum_k\!\left[\left(1\!\mp\!2z_{2}\!-\!z_{12}\right)\!\left(1\!\mp\!2z_{1}\!+\!z_{12}\right)\right]^{\frac{1}{2}}\!\sin[\frac{1}{2}(\theta_{1}\!-\!\theta_{2}\!\mp\!\theta_{12})],\nonumber
\end{eqnarray}
where the first sign that appears holds for $k=1 $, while the second for $k=2$. The equations of motion for the atom imbalances~(\ref{zs}) read
\begin{align}
\dot{z}_{k} & =-J\sqrt{(1\pm z_{12})^{2}-4 z_{k}^{2}}\sin\left(\theta_{k}\right)\label{eq:z_k}\\
 &\hspace{-0.40cm}\!-\! \frac{S_+}{2} \!\left[(1\!-\!2z_{2}\!-\!z_{12})(1\!+\!2z_{1}\!+\!z_{12})\right]^{\frac{1}{2}}\! \sin\left[\frac{1}{2}(\theta_{1}\!+\!\theta_{2}\!+\!\theta_{12})\right]\nonumber\\
 &\hspace{-0.40cm}\!-\! \frac{S_-}{2} \!\left[(1\!+\!2z_{2}\!-\!z_{12})(1\!-\!2z_{1}\!+\!z_{12})\right]^{\frac{1}{2}}\! \sin\left[\frac{1}{2}(\theta_{1}\!+\!\theta_{2}\!-\!\theta_{12})\right]\nonumber\\
 &\hspace{-0.40cm}\!\mp\!   \frac{1}{2}\bar \delta\!\left[(1\!+\!2z_{2}\!-\!z_{12})(1\!+\!2z_{1}\!+\!z_{12})\right]^{\frac{1}{2}}\! \cos\left[\frac{1}{2}(\theta_{1}\!-\!\theta_{2}\!+\!\theta_{12})\right]\nonumber\\
 &\hspace{-0.40cm}\!\pm\!   \frac{1}{2}\bar \delta\!\left[(1\!-\!2z_{2}\!-\!z_{12})(1\!-\!2z_{1}\!+\!z_{12})\right]^{\frac{1}{2}}\! \cos\left[\frac{1}{2}(\theta_{1}\!-\!\theta_{2}\!-\!\theta_{12})\right]\!,\nonumber
\end{align}

\begin{align}
\dot{z}_{12} & =\label{eq:z_12}\\
 &\hspace{-0.5cm}\!-\!    S_+\!\left[(1\!-\!2z_{2}\!-\!z_{12})(1\!+\!2z_{1}\!+\!z_{12})\right]^{\frac{1}{2}}\! \sin\!\!\left[\!\frac{1}{2}(\theta_{1}\!+\!\theta_{2}\!+\!\theta_{12})\!\right]\nonumber\\
 &\hspace{-0.5cm}\!+\!   S_-\!\left[(1\!+\!2z_{2}\!-\!z_{12})(1\!-\!2z_{1}\!+\!z_{12})\right]^{\frac{1}{2}}\! \sin\!\!\left[\!\frac{1}{2}(\theta_{1}\!+\!\theta_{2}\!-\!\theta_{12})\!\right]\nonumber\\
 &\hspace{-0.5cm}\!-\!   \bar \delta\!\sum_k\!\left[(1\!\pm\!2z_{2}\!-\!z_{12})(1\!\pm\!2z_{1}\!+\!z_{12})\right]^{\frac{1}{2}}\! \cos\!\!\left[\!\frac{1}{2}(\theta_{1}\!-\!\theta_{2}\!\pm\!\theta_{12})\!\right],\nonumber
\end{align}
with  the upper sign  for $k=1$ and the lower one for $k=2$. The equations for the phases~(\ref{thetas}) are
\begin{align}
\dot{\theta}_{k} &\! =\!N\left(U_k z_{k}\!+\!U_{12} z_{3-k}\right)+\frac{4J z_{k}\cos\left(\theta_{k}\right)}{\left[( z_{12}\!\pm\!1)^{2}\!-\!4 z_{k}^{2}\right]^{\frac{1}{2}}}\label{eq:theta_k}\\
 & + S_+\frac{\left[1\!-\!2z_{2}\!-\!z_{12}\right]^{\pm\frac{1}{2}}}{\left[1\!+\!2z_{1}\!+\!z_{12}\right]^{\pm\frac{1}{2}}}\cos[\frac{1}{2}(\theta_{1}\!+\!\theta_{2}\!+\!\theta_{12})]\nonumber \\
 & + S_-\frac{\left[1\!+\!2z_{2}\!-\!z_{12}\right]^{\pm\frac{1}{2}}}{\left[1\!-\!2z_{1}\!+\!z_{12}\right]^{\pm\frac{1}{2}}}\cos[\frac{1}{2}(\theta_{1}\!+\!\theta_{2}\!-\!\theta_{12})]\nonumber \\
 & +\bar \delta\frac{\left[1\!+\!2z_{2}\!-\!z_{12}\right]^{\pm\frac{1}{2}}}{\left[1\!+\!2z_{1}\!+\!z_{12}\right]^{\pm\frac{1}{2}}}\sin[\frac{1}{2}(\theta_{1}\!-\!\theta_{2}\!+\!\theta_{12})]\nonumber \\
 & +\bar \delta\frac{\left[1\!-\!2z_{2}\!-\!z_{12}\right]^{\pm\frac{1}{2}}}{\left[1\!-\!2z_{1}\!+\!z_{12}\right]^{\pm\frac{1}{2}}}\sin[\frac{1}{2}(\theta_{1}\!-\!\theta_{2}\!-\!\theta_{12})]\nonumber
\;,\end{align}

\begin{align}
\dot{\theta}_{12} & =4\Gamma\!+\!  N\left[\frac{U_{1}}{2} (z_{12}\!+\!1) \!+\! \frac{U_{2}}{2} (z_{12}\!-\!1) \!-\! U_{12}z_{12}\right]\nonumber\\
&\hspace{-0.5cm}\!-\! 2J\sum_k\frac{(z_{12}\pm 1)}{\sqrt{(1\pm z_{12})^{2}-4 z_{k}^{2}}}\cos\left(\theta_{k}\right) \label{eq:theta_12}\\
&\hspace{-0.5cm}\!+\! 2S_+\!\frac{z_{1}+z_{2}+ z_{12}}{\left[1\!-\!2z_{2}\!-\!z_{12}\right]^{\frac{1}{2}}\left[1\!+\!2z_{1}\!+\!z_{12}\right]^{\frac{1}{2}}}\cos[\frac{1}{2}(\theta_{1}\!+\!\theta_{2}\!+\!\theta_{12})]\nonumber \\
&\hspace{-0.5cm}\!+\!  2S_-\!\frac{z_{1}+z_{2}- z_{12}}{\left[1\!+\!2z_{2}\!-\!z_{12}\right]^{\frac{1}{2}}\left[1\!-\!2z_{1}\!+\!z_{12}\right]^{\frac{1}{2}}}\cos[\frac{1}{2}(\theta_{1}\!+\!\theta_{2}\!-\!\theta_{12})]\nonumber \\
&\hspace{-0.5cm}\!-\! 2\bar \delta\sum_k\!\!\frac{z_{1}-z_{2}\pm z_{12}}{\left[1\!\pm\!2z_{2}\!-\!z_{12}\right]^{\frac{1}{2}}\left[1\!\pm\!2z_{1}\!+\!z_{12}\right]^{\frac{1}{2}}}\!\sin[\frac{1}{2}(\theta_{1}\!-\!\theta_{2}\!\pm\!\theta_{12})],\nonumber
\end{align}
where, again, the upper(lower) sign corresponds to $k=1(2)$. We have reduced the problem from 8 to 6
equations and checked that these equations give the same numerical results than Eqs.~(\ref{eq:Nkalpha1})-(\ref{eq:thekalpha1}). Notice that if $ S_\pm$, $\Gamma$, and $\bar \delta$ vanish,
these equations give back those of the two-component two-well problem
discussed in Ref.~\cite{brunonjp}. According to its definition,
Eq.~(\ref{zs}), the polarization $z_{12}$ is bounded to the interval
$[1,-1]$. The two extremes of this interval correspond to all
atoms fully polarized on either internal state $1$ or $2$, respectively.
For each value of $z_{12}$ it is easy to show that the population
imbalance in each component is bounded by $|z_k|=(1\pm z_{12})/2 $,
where the minus sign corresponds to $k=2$.

\section{Josephson dynamics in the Spin-Orbit coupled double well}
\label{s5}

\begin{figure*}
\includegraphics[width=1.6\columnwidth]{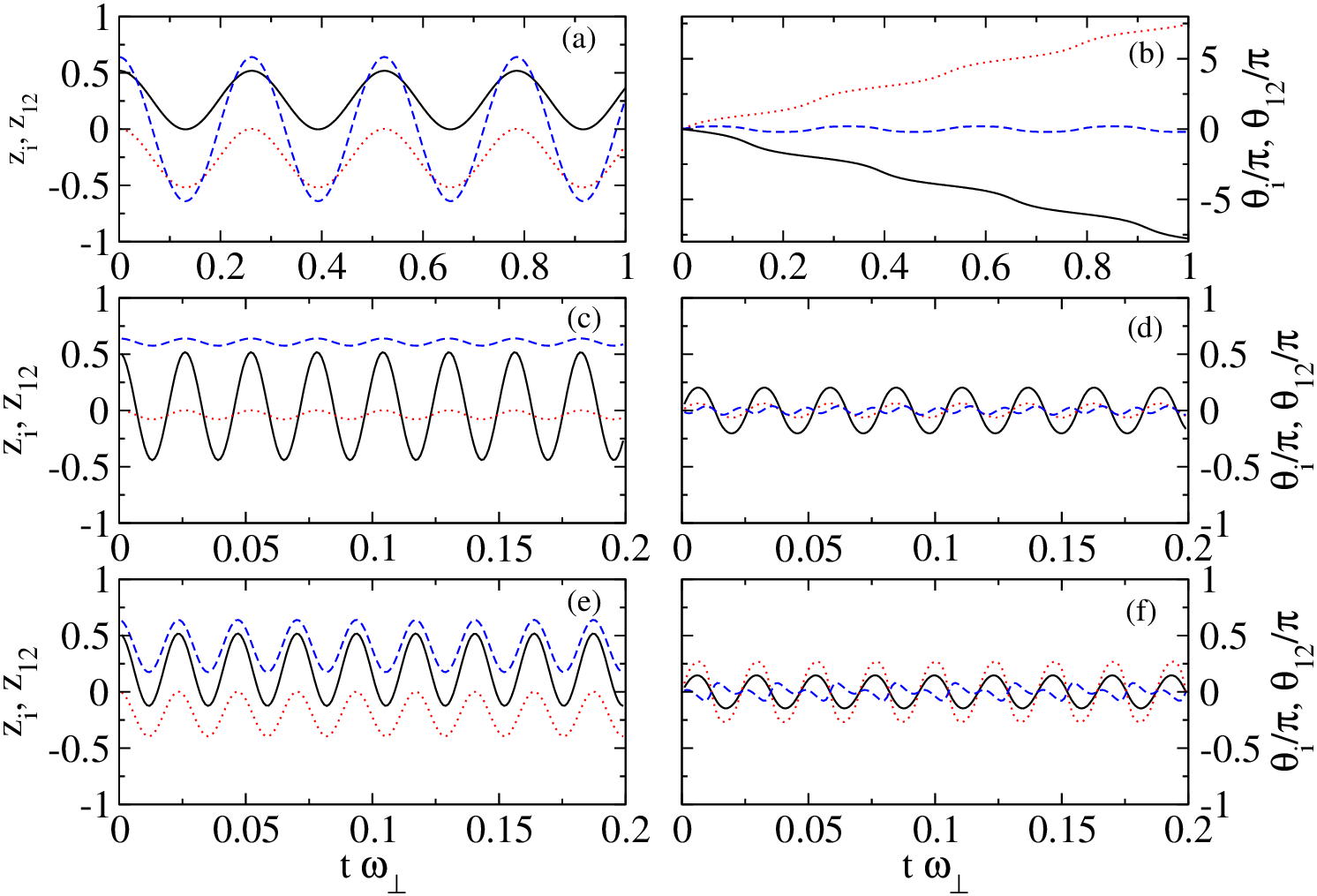}
\caption{(Color online) { {\it Macroscopic quantum
tunneling in the presence of a SO term.}
(a)  population imbalances  $z_1$ (solid black line),
$z_2$ (dotted red line), and polarization $z_{12}$ (dashed blue line) when $\Lambda_S=10$ in the absence of tunneling ($J=0$)).
(b) corresponding phases. (c) and (d) population imbalances, polarization, and
phases when $\Lambda_S =10$ in the presence of tunneling ($\Lambda_J =1$). (e) and (f) same when $\Lambda_S =2$. In all cases, $U=U_{12}=\tilde \delta=\Gamma=0$.Initial conditions: $z_1(0)=0.518$, $z_2(0)=0.002$, $z_{12}=0$, $\theta_k(0)=0$ ($k=1,2$), $\theta_{12}(0)=0$.
\vspace{-.01cm}} \label{fig1}}
\end{figure*}

For the numerical results discussed in this section, we consider
$N=10^5$, $^{87}$Rb atoms, and that the wavelength of the two
counter-propagating lasers is $\lambda\approx 10^3$ nm. Following
Ref.~\cite{so-bose1}, we introduce natural units for the momentum
and energy as $k_L=\sqrt{2}\pi/\lambda$ and $E_L=\hbar^2k_L^2/(2m)$.
The SO coupling is given by $v=E_L/\hbar k_L$ and therefore
for $\omega_\perp=400 \pi $Hz one obtains $\gamma=3.37$. Notice
that $S_\pm$ is defined as an overlap integral (see Eq.~(\ref{parameters})).
To keep the model simple we approximate the four on-site modes by
Gaussian wave functions. In this case, the overlap integral
is proportional to $\exp(-d^2)$, $d$ being the distance between
the minimum of each well and the origin. Then,  both $S_\pm$ and $J$ can be tuned by varying
the distance between the wells.

The SO coupling $v$ is independent on the detuning
$\delta$ and the Rabi coupling $\Omega_R$. We assume that $\Omega_R$
can be tuned in the interval $\left[0,7E_L/\hbar\right]$, and
then $\Gamma_{\mathrm{max}}=19.92$. $\bar \delta$ is proportional to
$\exp(-d^2)$ and can thus also be tuned by varying $d$.
In the following, we take the scattering lengths $a_1\simeq a_2\simeq 101.8 a_0$~\cite{mariona}, with
$a_0$ the Bohr atom radius which gives $U_{\mbox{ref}} \approx0.0012$, where $U_{\mbox{ref}}=\displaystyle {\frac{2a_{00}}{a_{\perp}} \int dx (\phi^{\alpha}(x))^4}$, 
$a_{00}=101.8a_0$, and the integral performed on the whole real axis. We take this value of the interactions to refer all variables in the rest of the paper.
We define $\Lambda_J\equiv U_{\mbox{ref}} N/J$, which is the usual variable
quantifying the ratio between atom-atom interaction and
tunneling in a single component bosonic Josephson junction. Similarly, we define the quantities $\Lambda_S \equiv U_{\mbox{ref}} N/S_+$, and  $\Lambda_D \equiv U_{\mbox{ref}} N/\tilde\delta$. Finally, we assume that the interactions can be tuned with respect to the reference, and therefore we define $C_U\equiv U/U_{\mbox{ref}}$ and $C_{U_{12}}\equiv U_{12}/U_{\mbox{ref}}$.

Note that in this part of the paper, we shall take $U_{12}$=0 to make the
effect of the spin-orbit coupling as clear and neat as possible. Let us note, however, that the effect of the repulsion between
species has been discussed thoroughly in Refs.~\cite{lobo,xu,satja,diaz1,ajj1,ajj2,qiu,rabijosephson,brunonjp},
where it was found that the conventional
Josephson dynamics is crucially modified by this term,
even leading to measure synchronization in some
limits~\cite{qiu13}. Particularly, the inter-species interaction induces the
presence of new fixed points on the problem which are
related to the repulsion between components. Nevertheless, we take into account the effect of the boson-boson repulsion in subsection~\ref{sec:interspec}.

\subsection{Some considerations about fixed points}
\label{sec:fixpoints}

Inspection of Hamiltonian~(\ref{semiclassicalH}) (or its many-body counterpart, Eqs.~(\ref{eq:two-level}), (\ref{eq:H_JU}), (\ref{eq:Hintersepcies})) permits one to identify three different processes which interchange atoms between wells or components. The first one is the usual tunneling between both wells, and is given by the term $J\sqrt{N_{k}^\alpha N_{k}^{\alpha'}}
\cos\left(\theta_{k}^\alpha-\theta_{k}^{\alpha'}\right)$ in Eq.~(\ref{semiclassicalH}). The second one is associated to the SO terms proportional to $S_\pm$ in Eq.~(\ref{semiclassicalH}), and couples the atoms
of one species located in one well to the atoms of the second species located in the other well. The third one
is given by  the terms corresponding to the detuning $\tilde \delta$ in Eq.~(\ref{semiclassicalH}). This is a coupling between atoms of different species in the same well. Finally, we note that the Rabi frequency $\Gamma$  introduces an energy gap between both species in Eq.~(\ref{semiclassicalH}). In this work, we study the effect of the SO coupling, detuning, and Rabi frequencies  in the well-known Josephson dynamics in double wells. We focus in the case in which the most populated species has certain population imbalance, and study the effect of the dynamics of this species on the second initially balanced species. To understand this problem, let us discuss briefly the fixed points of Eqs.~(\ref{eq:z_k})-(\ref{eq:theta_12}) when SO coupling and detuning frequencies are considered.

In the absence of interactions and when all terms other than the tunneling energy vanish, the fixed points  are the usual ones at $(z_k^0,\theta_k^0)=(0,n\pi)$, $n\in\mathbb{Z}$. In such a case, there is no  process that can produce interchange of atoms between the two components. Therefore $z_{12}$ remains constant at its initial value.  This picture changes when the other two processes are  considered. In the presence of the SO coupling and tunneling, when all other terms vanish, one can prove that Eqs.~(\ref{eq:z_k})-(\ref{eq:theta_12}) vanish for $(z_k^0,\theta_k^0)=(z_{12}^0,\theta_{12}^0)=(0,0)$. When initially $z_{12}$ is different from zero and all other variables vanish,  the population imbalances $z_k$ and the phase $\theta_{12}$ will remain in their initial value,  while the equations of motion can be reduced to
\begin{eqnarray}
\dot z_{12}=-2S_+\sin\bar\theta\sqrt{1-z_{12}^2},\label{eq:zsimplS}\\
\dot{\bar\theta}=2S_+\cos\bar\theta\frac{z_{12}}{\sqrt{1-z_{12}^2}},
\end{eqnarray}
with $\bar\theta=(\theta_1+\theta_2)/2$. Therefore, both $z_{12}$ and $\bar \theta$ will oscillate during the evolution. In such a case, we observe numerically that $\theta_1$ and $\theta_2$  grow unbounded, with opposed sign. In addition, when initially all variables are zero except for $z_k$, the polarization  $z_{12}$ remains at its initial value at zero. On the other hand, $(\theta_1+\theta_2)/2$ and $z_1+z_2$ will oscillate, with the latter bounded by $\pm(z_1(0)+z_2(0))$, while $\theta_{12}$ also oscillates. We observe numerically that $z_1-z_2=z_1(0)-z_2(0)$ along evolution.

In the presence only of detuning,  Eqs.~(\ref{eq:z_k})-(\ref{eq:theta_12}) vanish for  $(z_k^0,\theta_k^0)=(0,0)$, but now it is necessary that  $(z_{12}^0,\theta_{12}^0)=(0,\pi)$.  When initially $z_{12}$ is different from zero, $\theta_{12}=\pi$, and all other variables vanish,  the population imbalances $z_k$ and the phase $\theta_{12}$ will remain in their initial value,  while the equations of motion can be reduced to
\begin{eqnarray}
\dot z_{12}=-2\tilde \delta\cos\tilde\theta\sqrt{1-z_{12}^2},\label{eq:zsimpld}\\
\dot{\tilde\theta}=-4\tilde \delta\sin\tilde\theta\frac{z_{12}}{\sqrt{1-z_{12}^2}},
\end{eqnarray}
with $\tilde\theta=(\theta_1-\theta_2)/2$. Therefore, both $z_{12}$ and $\tilde \theta$ will oscillate during the evolution. In case $z_k$ are different from zero initially, while $z_{12}$ is zero,  both $(\theta_1-\theta_2)/2$ and $z_1-z_2$ will oscillate, with the latter bounded by $\pm(z_1(0)-z_2(0))$. Now,  $z_{12}$ remains at its initial value, $\theta_{12}$ also oscillates, and we observe numerically that $z_1+z_2=z_1(0)+z_2(0)$.


In the next section we study how the fixed point analysis briefly discussed above can help to understand the dynamics when the most populated species has certain population imbalance, while the second species is balanced, that is initially $z_{12}$ and $z_1$ are non-zero, while $z_2$ is zero. To illustrate this situation we consider in all numerical examples to be discussed in the next subsections  that initially $z_1=0.518$, $z_2=0.002$, $z_{12}=0.64$ and all initial phases are zero unless explicitly indicated.

\subsection{Macroscopic quantum tunneling and self-trapping
in the presence of Spin-Orbit coupling}

We first assume $\bar \delta=\Gamma=0$, and study the effect of
the SO coupling $v$. This coupling  is associated to
the kinetic moment $p_x$ or $p_y$ of the atoms in each species
(see Eq.~(\ref{eq:SP_part})). As we have shown, when the single
particle 3D potential can be reduced effectively to a 1D
double-well, where the dynamics in transverse directions is
essentially frozen, the SO coupling, proportional to $S_\pm$, becomes
apparent in a non-trivial way in the equations of motion.

The SO coupling term allows for the complete transfer of the atoms of species 1 in the left well to species 2 in the right well (see Fig.\ref{fig1}a-b), when no other term is considered. 
In the presence of a tunneling term which dominates the SO coupling, see Figs.~\ref{fig1}c-d, $z_1$ shows fast Rabi oscillations, where its corresponding phase is bounded. This can be understood in view of Eqs.~(\ref{eq:z_k}) and~(\ref{eq:theta_k}), as in the presence of a tunneling which dominates over $S_\pm$, these equations will only vanish if $\theta_k=0,\pi,\dots$
Therefore, in this case $\theta_k$ cannot grow unbounded and has to oscillate around $\theta_k=0$.  A small transfer of atoms between both components still occurs, but it is not enough to transfer all population from  component 1  to component 2 before it tunnels to the other well. If $S_\pm$ are comparable to $J$, both effects are combined, and the transfer of atoms between components  is enlarged and the tunneling of the  atoms of component 1 is reduced, as shown in  Figs.~\ref{fig1}e-f.

\begin{figure}[t]
\includegraphics[width=\columnwidth]{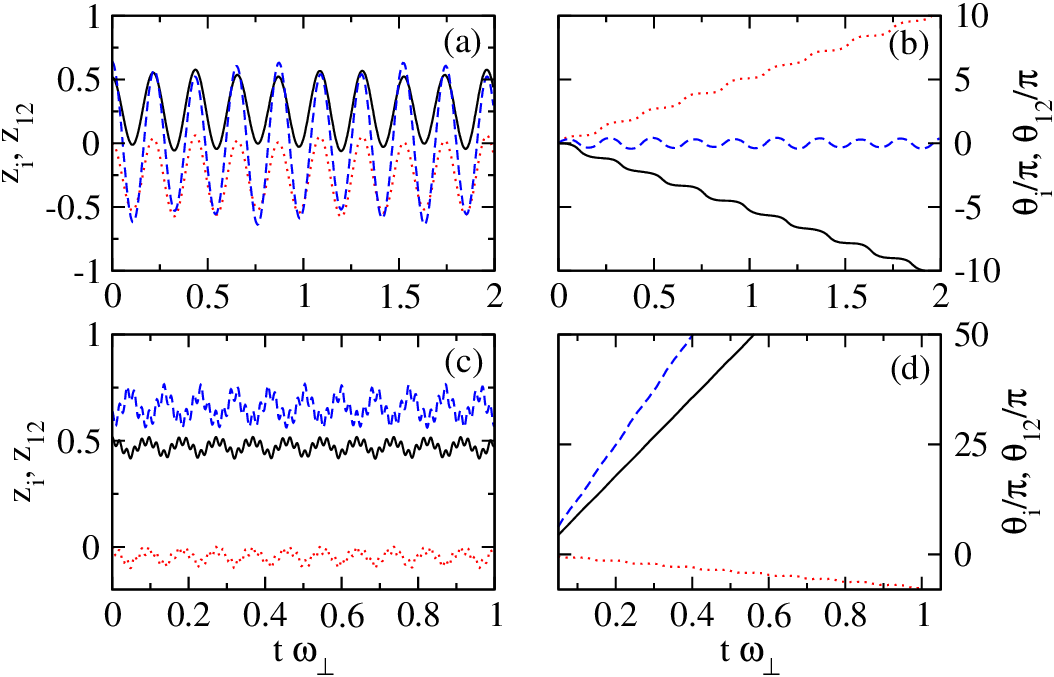}
\caption{(Color online) {\it Macroscopic
quantum self-trapping  in the presence of a SO term.}
(a)  population imbalances  $z_1$,
$z_2$, and polarization $z_{12}$ when $\Lambda_S=10$ when the interactions are small, $C_U=1/5$. Same color conventions as in Fig.~\ref{fig1}.
(b) corresponding phases. (c) and (d) self-trapping induced by the interactions ($\Lambda_S=10,\,C_U=5$). In all cases, $U_{12}=J=\tilde \delta=\Gamma=0$.
Initial conditions as in Fig.~\ref{fig1}.
\vspace{-.01cm}} \label{fig2}
\end{figure}

\begin{figure}[t]
\includegraphics[width=\columnwidth]{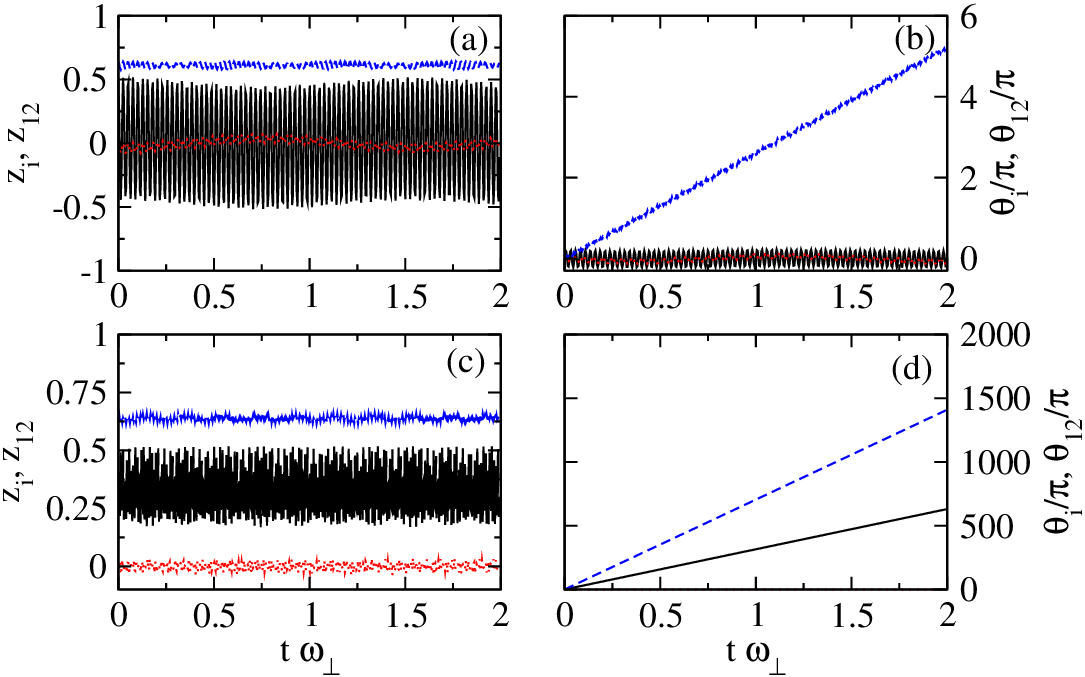}
\caption{(Color online) {\it Macroscopic
quantum tunneling and self-trapping  in the presence of a SO term.}
(a)  population imbalances  $z_1$,
$z_2$, and polarization $z_{12}$  for $\Lambda_S=10$ when the interactions  are small ($C_U=1/10$) in the presence of tunneling, $\Lambda_J=1$, showing self-trapping dynamics of $z_{12}$. Same color conventions as in Fig.~\ref{fig1}.
(b) corresponding phases. (c) and (d) self-trapping dynamics in all variables  when $\Lambda_S=10$ and $C_U=25$.  In all cases, $U_{12}=\tilde \delta=\Gamma=0$.
Initial conditions as in Fig.~\ref{fig1}.
\vspace{-.01cm}} \label{fig3}
\end{figure}

Let us now illustrate the effect of the interactions on the dynamics induced by the SO coupling shown in Figs.~\ref{fig1}a-b corresponding to the absence of hopping and interactions. In the presence of a small interaction term, the transfer of atoms between components associated with  the dynamical evolution of the polarization $z_{12}$ still occurs. The interactions only modulate slightly this dynamics, as shown in Figs.~\ref{fig2}a-b. Conversely, for larger $U$, self-trapping occurs in all variables, and correspondingly all phases are running (see Figs.~\ref{fig2}c-d). When both tunneling and interactions are different from zero, self-trapping can occur independently on the polarization, $z_{12}$, and the population imbalances, $z_k$. In Figs.~\ref{fig3}a-b we show the dynamics when  the interactions dominate over the SO term, but not over the tunneling, which induces 
self-trapping  on $z_{12}$, while $z_k$ still oscillates. Increasing further the interactions produces self-trapping also in $z_k$, as plotted in  Figs.~\ref{fig3}c-d. We conclude this subsection with a remark about Figs.~\ref{fig1}a-b. The phase $(\theta_{1}+\theta_2)/2$ oscillates accordingly with $\theta_{1}$ and $\theta_{2}$ growing unbounded with opposed sign  (see Eq.~(\ref{eq:zsimplS})). Moreover, because $z_k$ are different from zero initially, $z_1+z_2$ oscillates, while $z_1-z_2=z_1(0)-z_2(0)=0.516$ along all the evolution, and $\theta_{12}$ oscillates,  in accordance with the discussion on section~\ref{sec:fixpoints}.

\subsection{Macroscopic quantum tunneling and self-trapping
in the presence of the Rabi and detuning frequencies}

\begin{figure*}
\includegraphics[width=1.6\columnwidth]{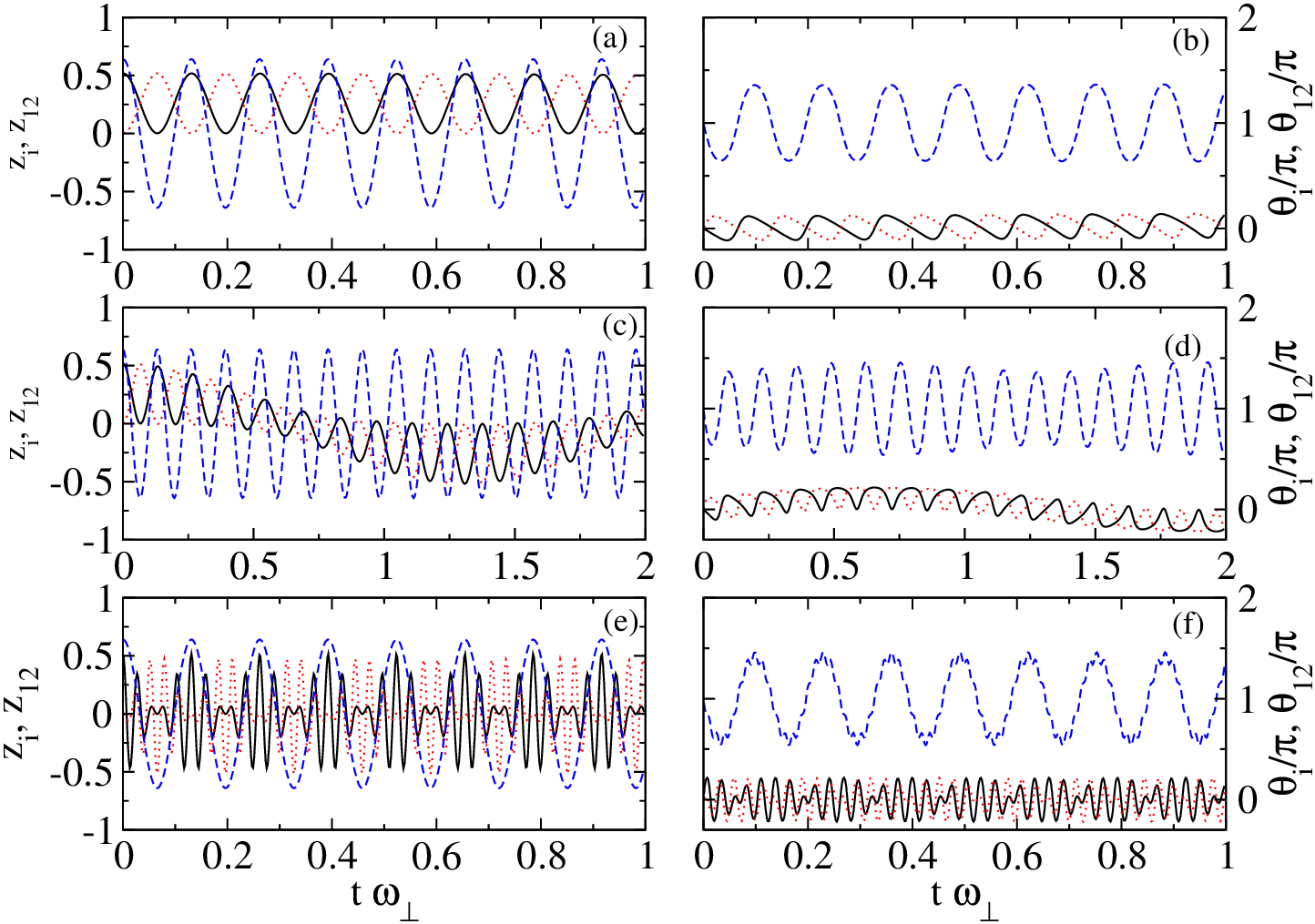}
\caption{(Color online) {\it Macroscopic
quantum tunneling  in the presence of a detuning $\tilde \delta$, when the initial phase for the polarization is $\theta_{12}(0)=\pi$.}
(a)  population imbalances  $z_1$,
$z_2$, and polarization $z_{12}$  for $\Lambda_D=5$ when the tunneling term is very small $ \Lambda_J=10^{3}$. Same color conventions as in Fig.~\ref{fig1}.
(b) corresponding phases. (c) and (d) same when $\Lambda_J=10^{2}$, and  (e) and (f)  when  $\Lambda_J=1$.  In all cases, $U=U_{12}=S_\pm=\Gamma=0$.
Initial conditions for $z_k$, $\theta_k$ ($k=1,2$), and $z_{12}$ as in Fig.~\ref{fig1}.
\vspace{-.01cm}} \label{fig4}
\end{figure*}

\begin{figure*}
\includegraphics[width=1.6\columnwidth]{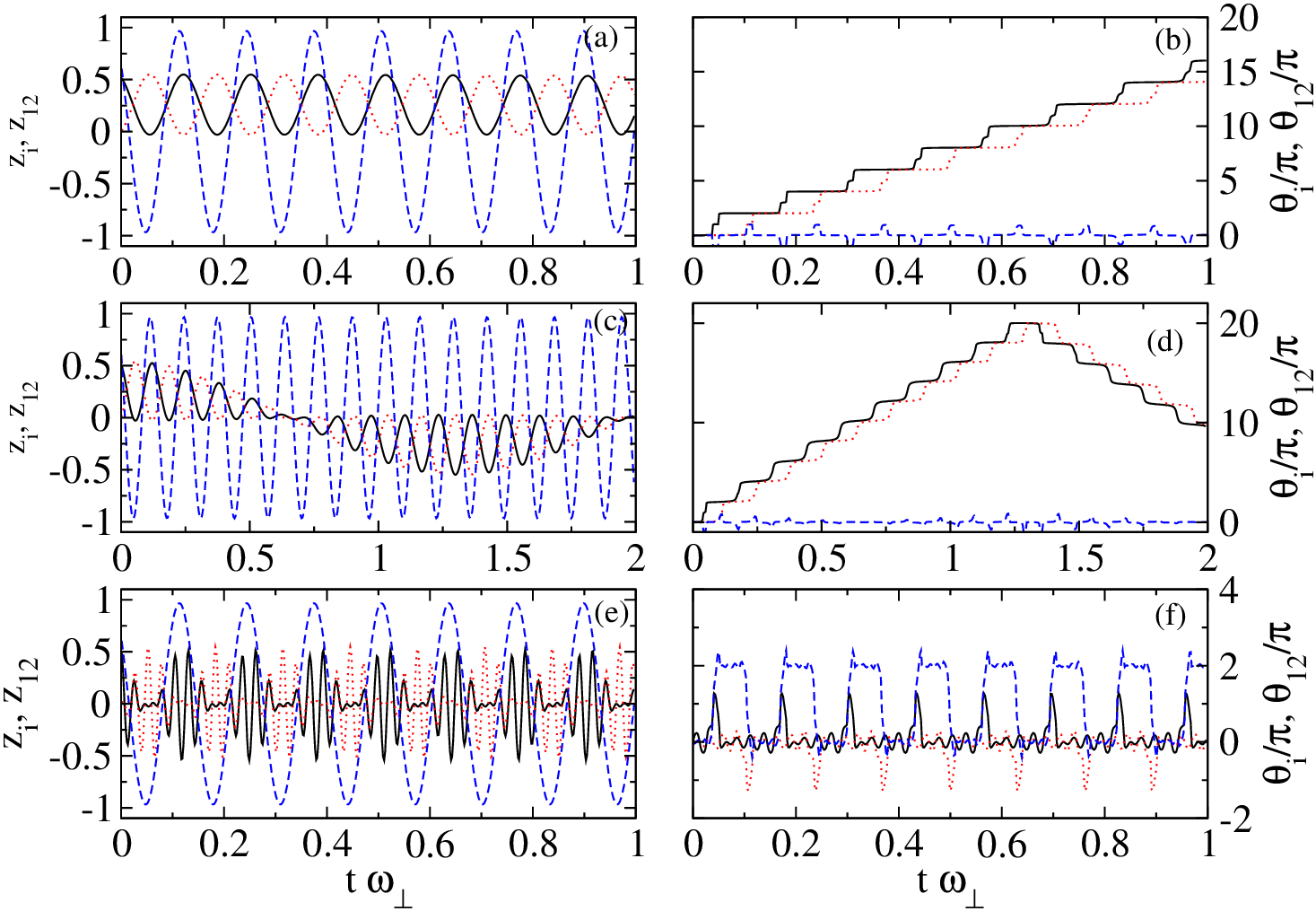}
\caption{(Color online) {\it Macroscopic
quantum tunneling in the presence of a detuning $\tilde \delta$, when the initial phase for the polarization is $\theta_{12}(0)=0$.}
(a)  population imbalances  $z_1$,
$z_2$, and polarization $z_{12}$  for $\Lambda_D=5$ when the tunneling term is very small $ \Lambda_J=10^{3}$. Same color conventions as in Fig.~\ref{fig1}.
(b) corresponding phases. (c) and (d) same when $\Lambda_J=10^{2}$, and  (e) and (f)  when  $\Lambda_J=1$.   In all cases, $U=U_{12}=S_\pm=\Gamma=0$.
Initial conditions $z_k$, $\theta_k$ ($k=1,2$), and $z_{12}$ as in Fig.~\ref{fig1}.
\vspace{-.01cm}} \label{fig5}
\end{figure*}

\begin{figure}
\includegraphics[width=\columnwidth]{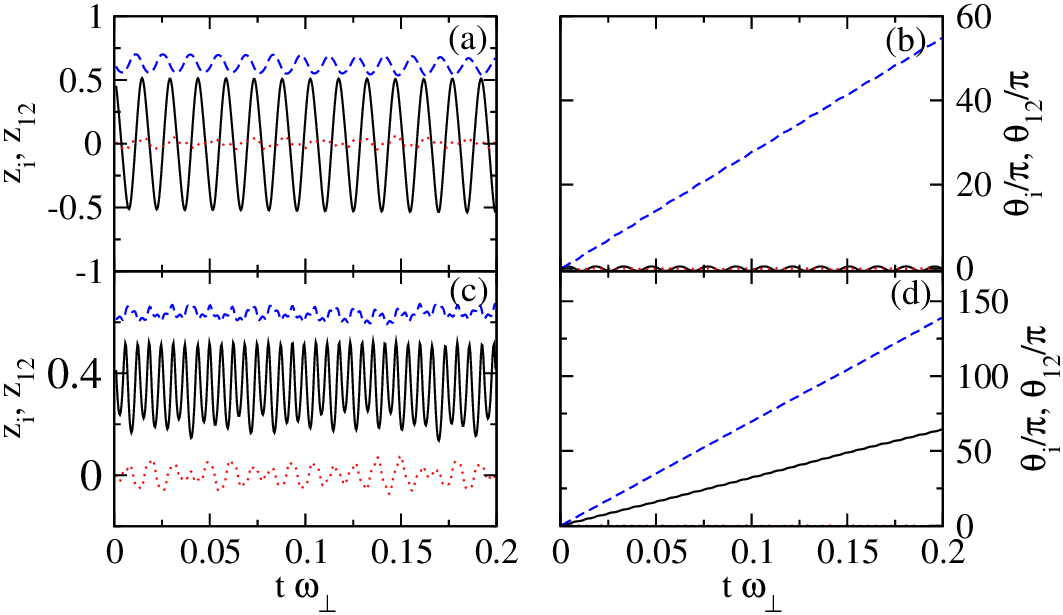}
\caption{{\it Macroscopic quantum self-trapping in the presence of a detuning $\tilde \delta$, when initially $\theta_{12}(0)=0$.} (a)  population imbalances  $z_1$,
$z_2$, and polarization $z_{12}$   when $\Lambda_D=5$,  $ \Lambda_J=1$, $C_U=10$, and all other terms vanish, showing self-trapping dynamics of $z_{12}$. Same color conventions as in Fig.~\ref{fig1}. (b) corresponding phases. (c) and (d) self-trapping in all variables  when the interactions are increased to $C_U=25$.
Initial conditions for $z_k$, $\theta_k$ ($k=1,2$), and $z_{12}$ as in Fig.~\ref{fig1}.
\vspace{-.01cm}} \label{fig6}
\end{figure}

Let us now discuss on the effect of the Rabi and detuning frequencies, $\Gamma$ and $\bar \delta$,
respectively. The detuning frequency induces a local transfer of population between both components.
To illustrate this, we represent in Figs.~\ref{fig4}a-b  the dynamics when all other terms are
zero and the tunneling $J$ is very small, when initially $\theta_{12}=\pi$, and all other phases
vanish. According to Eq.~(\ref{eq:zsimpld}), the corresponding dynamics will be oscillatory around
the fixed point, with $\theta_1-\theta_2$ also oscillating. Because initially $z_1=0.518$ and $z_2=0.002$,
$z_1-z_2$ also oscillates with $z_1+z_2=z_1(0)+z_2(0)=0.52$ along all the evolution. Again, $\theta_{12}$
also oscillates, in accordance with the discussion in section~\ref{sec:fixpoints}.  For larger $J$, the
atoms of each species tunnel also to the other well in the same time scales, as shown in
Figs.~\ref{fig4}c-d. Differently from the dynamics in the presence of the SO term, $\theta_1$ and
$\theta_2$ do not grow unbounded in the absence of tunneling. Therefore, atoms can still be
transformed from component 1 to component 2 in the presence of large tunneling. This effect is
quicker if $J$ is increased further, as illustrated in Figs.~\ref{fig4}e-f. The dynamics of
$z_{12}$ is not affected by the Josephson physics. This is due to the fact that the detuning
term induces local population transfer, similarly to the transfer of populations among the
different Zeeman components in a spinor BEC~\cite{spinor}. In Fig.~\ref{fig5} we reproduce
the same cases when initially $\theta_{12}=0$. As this initial condition does not correspond to a
fixed point, the phase oscillates abruptly and $z_{12}$ oscillates in the interval $[-1,1]$, as
it possibly corresponds to an initial condition close to a separatrix. In Figs.~\ref{fig6}a-b
we show that the interactions can induce self-trapping in $z_{12}$ when the interactions dominate
over the detuning term. If increased further, Figs.~\ref{fig6}c-d, the interactions induce self-trapping
in all variables.

\begin{figure}[t]
\includegraphics[width=\columnwidth]{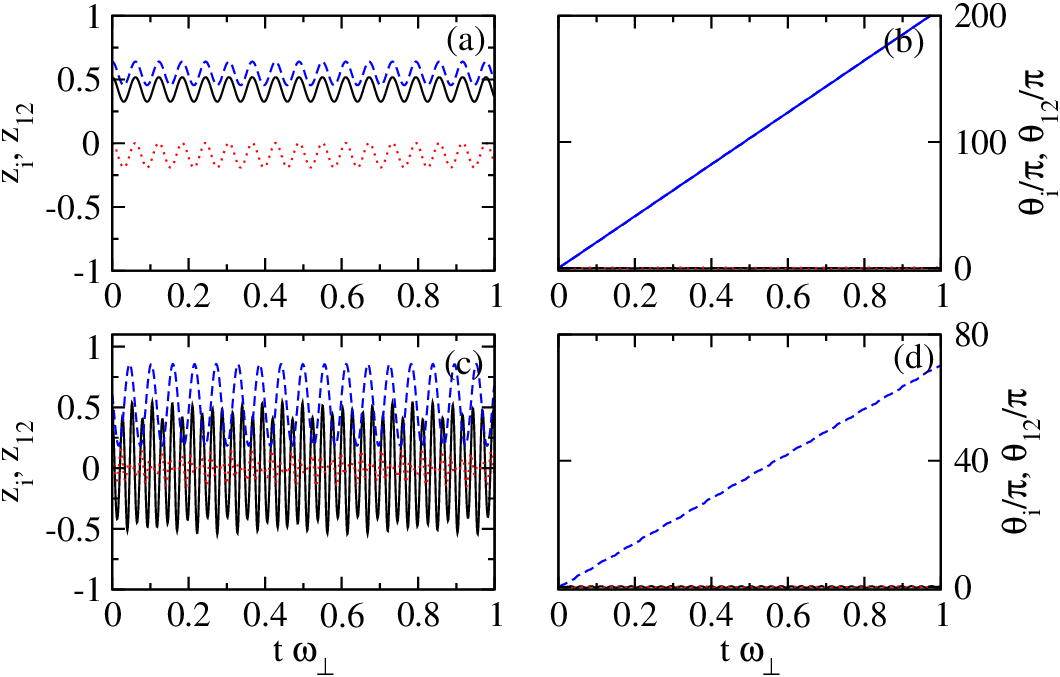}
\caption{{\it Decoupling induced by the Rabi frequency, $\Gamma$.}
(a)  population imbalances  $z_1$,
$z_2$, and polarization $z_{12}$   when  $\Lambda_S=10$, $\Gamma=50$, and all other terms vanish. Same color conventions as in Fig.~\ref{fig1}. (b) corresponding phases. (c) and (d) same  when $\Lambda_D=5$, $\Gamma=50$, $\Lambda_J=1$, and all other terms vanish.
Initial conditions as in Fig.~\ref{fig1}.
\vspace{-.01cm}} \label{fig7}
\end{figure}

The effect of the Rabi frequency  on the dynamics associated with  $S_\pm$ and $\delta$ can be understood in view
of the equations of motion for $\theta_{12}$,~Eq.~(\ref{eq:theta_12}), as $\Gamma$ appears as a constant in the equation.
Therefore, it introduces an energy gap between both components, which, when it dominates over the rest of terms,
forces $\theta_{12}$ to be a running phase, similarly to the problem of bosons in an excited level in
double wells~\cite{2014Gilletarxiv}. In Figs.~\ref{fig7}a-b we show the effect of $\Gamma$ in the
oscillations of the polarization $z_{12}$ due to the effect of the SO coupling $S_\pm$. We observe that the
oscillation of the polarization is reduced with respect to the case of Figs.~\ref{fig1}a-b, and $\theta_{12}$ is a running phase. Then, the effect of
$\Gamma$ is to inhibit the coupling of the two components associated to the SO term. Similarly,
in Figs.~\ref{fig7}c-d  we observe that $\Gamma$ again reduces (with respect to the case of Figs.~\ref{fig5}e-f) the oscillations on $z_{12}$ induced by 
the detuning frequency $\delta$, again decoupling the dynamics of the two components.

\subsection{Effect of the inter-species interaction}
\label{sec:interspec}

\begin{figure*}[t]
\includegraphics[width=1.6\columnwidth]{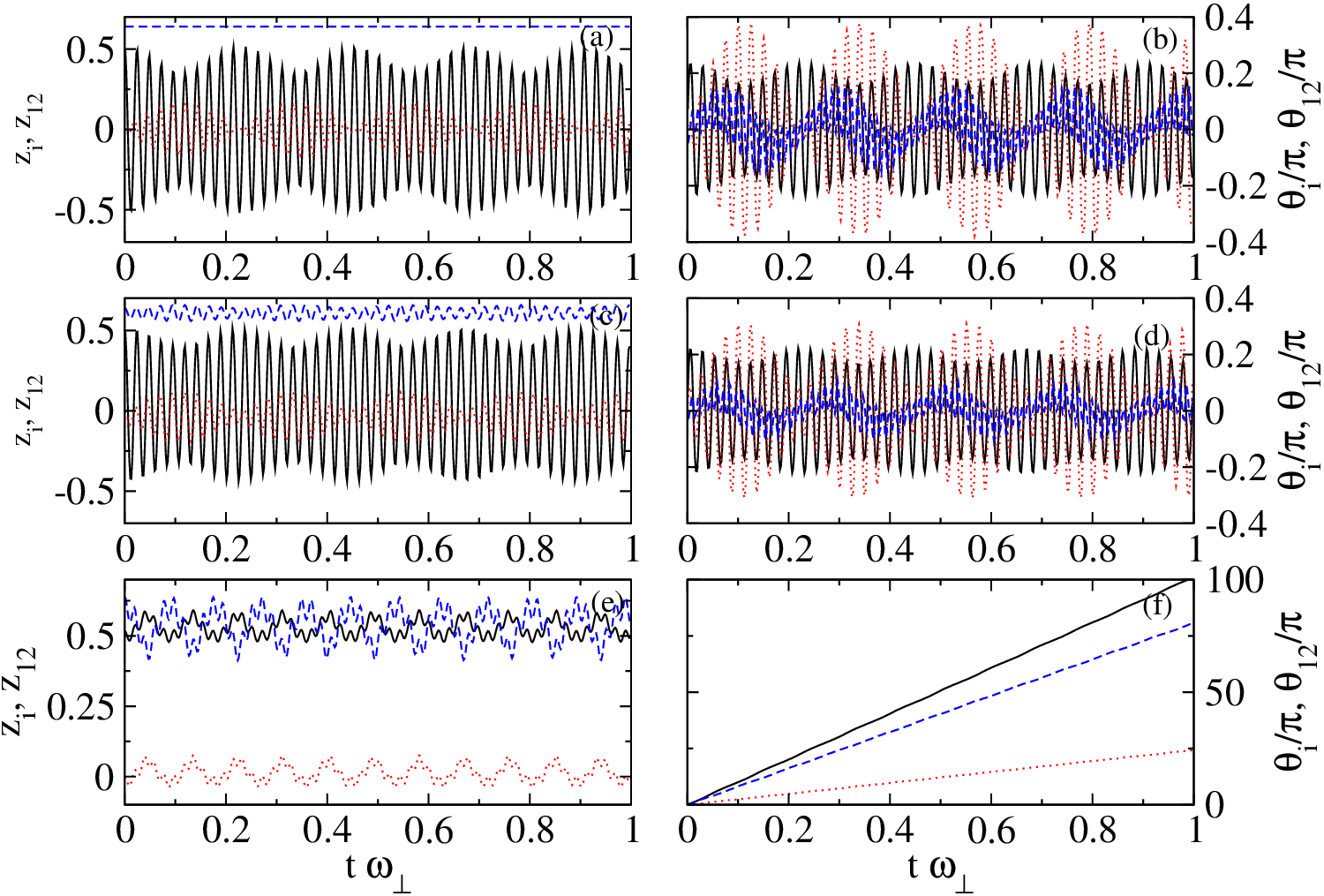}
\caption{{\it Macroscopic quantum tunneling and self-trapping in the presence of SO coupling and inter-species interactions  $U_{12}$.}
(a)  population imbalances  $z_1$,
$z_2$, and polarization $z_{12}$  when  $C_U=C_{U_{12}}=1/2 $ and $\Lambda_J=1$, and all other terms vanish, reproducing two-component dynamics in the absence of SO coupling. Same color conventions as in Fig.~\ref{fig1}. (b) corresponding phases.  (c) and (d) effect of the SO coupling term when   $\Lambda_S=10$.  (e), (f) dragging of $z_2$ by $z_1$ in the self-trapping dynamics, when   $\Lambda_S=10$, $C_U=5 $, $C_{U_{12}}=1 $, and all other terms vanish.
Initial conditions as in Fig.~\ref{fig1}.
\vspace{-.01cm}} \label{fig8}
\end{figure*}

As commented before, in the previous results we have decided to set the inter-species
interaction $U_{12}$ to zero, to emphasize the effects of the
spin-orbit coupling, Rabi and detuning frequencies. In a possible experimental realization
along the lines of the recent SO experiments~\cite{so-bose1} it
may be difficult to experimentally achieve this limiting case.
For the case of considering two of the Zeeman states of the
$F=1$ $^{87}$Rb one has that $U_1\approx U_2\approx U_{12}$ \cite{brunonjp}.
The  inter-species  interactions has profound and diverse effects on the Josephson dynamics of two components in double wells~\cite{lobo,xu,satja,diaz1,ajj1,ajj2,qiu,rabijosephson,brunonjp,2014Gilletarxiv}, and therefore an extensive discussion on this topic is out of the scope of this paper.
If the inter-species interaction $U_{12}$ is similar to
intra-species ones ($U_{1}$ and $U_{2}$), and for the case considered
here of a polarized initial state, the dynamics of the less populated
species is crucially influenced by the dynamics of the most populated
one~\cite{diaz1}. Moreover, for a value of $U_{12}$  above certain threshold, both species cover the same region in the phase portrait, an effect known as {\it measurement synchronization} (MS)~\cite{qiu,qiu13}. In Figs.~\ref{fig8}a-b we show the dynamics of the system when all the detunings and the SO coupling term vanish. Here, the value of $U_{12}$ is below the MS  threshold, as can be seen from the fact that $z_1$ and $z_2$ have different maximum amplitudes. In Figs.~\ref{fig8}c-d we have increased slightly the SO term, and some transfer of atoms between both species occurs. If the interactions dominate over the tunneling and the SO term (for example, as  in Figs.~\ref{fig2}c-d), all variables are self-trapped and the corresponding phases grow. In Figs.~\ref{fig6}g-h Figs.~\ref{fig8}e-f we plot the same case as in Figs.~\ref{fig2}c-d for non-zero inter-
species interactions. Moreover, the evolution of $z_2$ is slightly dragged by that of $z_1$, an  effect which is in accordance with the results of Ref.~\cite{diaz1}. We have also observed numerically that the phenomena associated with the detuning and Rabi frequencies described above still occur in the presence of the inter-species energy.

As a conclusive remark, we note that working within the single-particle Hamiltonian framework corresponding to Eq. (\ref{eq:SP_part}) produces changes with respect to the results achieved by Zhang {\it et al.} \cite{zhang}. In fact in our case, the spin-orbit
coupling induces change in polarization (the external josephson is transferred into internal) while in the case of \cite{zhang} it affects the tunneling rate.

\section{Summary and conclusions}
\label{s6}

The recent developments in ultracold atomic gases, namely the
experimental realization of external bosonic Josephson junctions,
together with the artificial creation of spin-orbit coupling
for ultracold atoms paved the way to discuss the interplay between both
effects in a common set-up. As we have described, the conventional
macroscopic quantum tunneling or self-trapping scenarios of two bosonic
components confined in a double-well determine crucially the
polarization induced by the spin-orbit (SO) coupling. We have shown that the
SO coupling transfers atoms between both components in different wells.
This population transfer induced by the SO term depends on how this energy compares
with the tunneling and interaction energy. In the macroscopic quantum tunneling
limit this transfer is large if it dominates over the tunneling energy. By
increasing the interactions we observe that one can induce the self-trapping
of the polarization variable and subsequently of the population imbalance.
Secondly, we have studied the effect of the Rabi and detuning frequencies.
We have shown that the effect of the Rabi frequency is to decouple both
components, and thus reduce the population transfer between the two species. On the
contrary, the effect of the detuning frequency is a coupling between both
components in the same well. Now, this transfer occurs even in the presence of
large tunneling energy, though in different time scales. The interactions produce
also self-trapping in all variables if they dominate over the tunneling and detuning.
Finally, we have shown  that  the phenomena  associated  with the SO coupling,
detuning and Rabi frequencies are robust in the presence of inter-species interactions.

\acknowledgments
This work has been supported by MIUR (PRIN 2010LLKJBX). LD, GM and LS acknowledge
financial support from the University of Padova (Progetto di Ateneo 2011) and
Cariparo Foundation (Progetto di Eccellenza 2011). GM acknowledges financial
support also from Progetto Giovani 2011 of University of Padova. LD acknowledges
financial support also from MIUR (FIRB 2012 RBFR12NLNA). This work has been
supported by Grants No. FIS2011-24154 and No. 2009-SGR1289. B.J.D. is supported by the Ram\'on y Cajal program.

\appendix
\section{Many-Body Hamiltonian for the spin-orbit effect in double-wells}
\label{sec:appendix1}

The second quantized Hamiltonian for $N$ interacting two-component
bosons of equal mass $m$ confined by an external potential $U(\mathbf{r})$ in terms of the creation-annihilation
field operators - $\hat{\Psi}(\mathbf{r})$, $\hat{\Psi}^{\dagger}(\mathbf{r})$ -
in the presence of spin-orbit coupling is
\begin{align}
\hat{H} & =  \int d\mathbf{r}\hat{\Psi}^{\dagger}(\mathbf{r})
H_{\mathrm{sp}}\hat{\Psi}(\mathbf{r})\label{eq:second-quantized1}\\
\hspace{-1.5cm} & +  \frac{1}{2}\int d\mathbf{r}
\hat{\Psi}^{\dagger}(\mathbf{r})\left[\int d\mathbf{x'}
\hat{\Psi}^{\dagger}(\mathbf{r}')
V_{int}(\mathbf{r}-\mathbf{r}')
\hat{\Psi}(\mathbf{r})\right]
\hat{\Psi}(\mathbf{r}),\nonumber
\end{align}
where $V_{int}(\mathbf{r}-\mathbf{r}')$ stands for the two-body
interaction, and $\hat{\Psi}$ and $\hat{\Psi}^{\dagger}$ are two-component vectors.
Here $H_{\mathrm{sp}}$ is
\begin{eqnarray}
H_{\mathrm{sp}} & = &
\left(-\frac{\hbar^{2}}{2m}\nabla^{2}+U(\mathbf{r})\right)\sigma_{0}
+v_{R}\left(p_{x}\sigma_{y}-p_{y}\sigma_{x}\right)\nonumber\\
 & + & v_{D}\left(p_{x}\sigma_{y}
+p_{y}\sigma_{x}\right)
+\frac{\Omega_{R}}{2}\sigma_{z}
+\frac{\delta}{2}\sigma_{y},
\end{eqnarray}
with $\sigma_{0}$ the $2\times2$ identity matrix, and $\sigma_{x,y,z}$
the Pauli matrices.
Let us write $\hat{\Psi}(\mathbf{r})=
\left(\hat{\Psi}_{1}(\mathbf{r}),\hat{\Psi}_{2}(\mathbf{r})\right)$,
where the index $1$ ($2$) accounts for the first (second) component.
We consider contact interactions both for intra-species interaction and for inter-species one. This means that the interatomic potential for the former case is assumed to be $g_k\delta(\mathbf{r}-\mathbf{r}')$ ($g_k=4\pi\hbar^2a_k/m$ with $a_k$ the intra-species $s$-wave scattering length), while for the latter $g_{12}\delta(\mathbf{r}-\mathbf{r}')$ ($g_{12}=4\pi\hbar^2a_{12}/m$ with $a_{12}$ the inter-species $s$-wave scattering length). Then, the interacting part
of the Hamiltonian can be written in the following way
\begin{eqnarray}
\hat{H}_{\mathrm{int}} &\!= \!& \sum_{k=1,2}
\frac{g_{k}}{2}
\!\int d\mathbf{r}\hat{\Psi}_{k}^{\dagger}
\hat{\Psi}_{k}^{\dagger}\hat{\Psi}_{k}
\hat{\Psi}_{k}+g_{12}
\!\int d\mathbf{r}
\hat{\Psi}_{1}^{\dagger}\hat{\Psi}_{2}^{\dagger}
\hat{\Psi}_{2}\hat{\Psi}_{1}.\nonumber
\end{eqnarray}
We can write the Hamiltonian as
$\hat{H}=\hat{H}_{1}+\hat{H}_{2}+\hat{H}_{12}$ with
\begin{eqnarray}
\hat{H}_{1} &\! =\! & \int d\mathbf{r}
\hat{\Psi}_{1}^{\dagger}H_{\mathrm{sp}}^{1}\hat{\Psi}_{1}
 + \frac{g_{1}}{2}\int d\mathbf{r}
\hat{\Psi}_{1}^{\dagger}\hat{\Psi}_{1}^{\dagger}\hat{\Psi}_{1}\hat{\Psi}_{1}
\end{eqnarray}
and similarly for species 2. Here, the single particle Hamiltonian
is $H_{\mathrm{sp}}^{k}=\displaystyle{\left(-\frac{\hbar^{2}}{2m}\nabla^{2}
+U(\mathbf{r})\pm\Omega_R/2\right)}$ with the plus or minus for
$1$ and $2$, respectively. The part of the Hamiltonian which couples both bosonic species is
\begin{align}
\hat{H}_{12} &\! = \!
   \hbar\int \!d\mathbf{r}\hat{\Psi}_{1}^{\dagger}\left[ v_{R}\left(-\partial_{x}+i\partial_{y}\right)+  v_{D}\left(-\partial_{x}-i\partial_{y}\right)\right]\hat{\Psi}_{2}\nonumber\\
 &\! +\! \hbar\int \!d\mathbf{r}\hat{\Psi}_{2}^{\dagger}\left[ v_{R}\left(\partial_{x}+i\partial_{y}\right)+ v_{D}\left(\partial_{x}-i\partial_{y}\right)\right]\hat{\Psi}_{1}\\
&\! + \!g_{12}\int \! d\mathbf{r}\hat{\Psi}_{1}^{\dagger}\hat{\Psi}_{2}^{\dagger}\hat{\Psi}_{2}\hat{\Psi}_{1}
\!-\!i\!\int \! d\mathbf{r}\hat{\Psi}_{1}^{\dagger}\frac{\delta}{2}\hat{\Psi}_{2}+\!i\!\int\! d\mathbf{r}\hat{\Psi}_{2}^{\dagger}\frac{\delta}{2}\hat{\Psi}_{1}
\;.\nonumber
\end{align}
We assume a separable potential, with harmonic confinement in the
$y$ and $z$ directions and a double-well in the $x$ direction
$U(\mathbf{r})=\frac{1}{2}m\omega_{\perp}^{2}(y^{2}+z^{2})+V_{DW}(x)$.
Let us write $\hat{\Psi}_{1}(\mathbf{r})=\hat{a}_{L}\psi_1^L(\mathbf{r})+\hat{a}_{R}\psi_1^R(\mathbf{r})$ and
$\hat{\Psi}_{2}(\mathbf{r})=\hat{b}_{L}\psi_{2}^{L}(\mathbf{r})+\hat{b}_{R}\psi_2^R(\mathbf{r})$, with $\hat{a}_{L(R)}$ and $\hat{b}_{L(R)}$ operators annihilating a boson of the species 1 and a boson of the species 2, respectively, in the left (right) well. These single-particle operators  obey the usual boson-commutation relations. For the orbitals $\psi_{k}^{\alpha}(\mathbf{r})$ we use Gaussian-like functions for the $y$ and $z$ directions (i.e., the ground-state wave function of the harmonic oscillator $m\omega_{\perp}^2y^2/2$ times that of the harmonic oscillator $m\omega_{\perp}^2z^2/2$) and
on-well localized functions $w_{k}^{\alpha}(x)$ ($\int dx w_{k}^{\alpha *} w_{k}^{\alpha'}=\delta_{\alpha,\alpha'}$) in the $x$ direction.
In such a way, we have that $\psi_{k}^{\alpha}(\mathbf{r})=\frac{1}{\sqrt{\pi}a_{\perp}}
\exp\left(-\frac{y^{2}+z^{2}}{2a_{\perp}^{2}}\right)w_k^\alpha(x)$ with
$a_{\perp}=\sqrt{\hbar/m\omega_{\perp}}$, $k=1,2$ and $\alpha=L,R$. Then, we obtain
\begin{equation}
\hat{H}=\hat{H}_{1}+\hat{H}_{2}+\hat{H}_{12}.\label{eq:two-level}
\end{equation}
The first two terms are
\begin{equation}
\hat{H}_{1}=(\hat{n}_{1}^L+\hat{n}_1^R)\varepsilon_{1}-J_{1}\sum_{\alpha\ne \alpha'}\hat{a}_{\alpha}^{\dagger}\hat{a}_{\alpha'}+\frac{U_{1}}{2}\sum_{\alpha}\hat{n}_1^\alpha\left(\hat{n}_1^\alpha-1\right),\label{eq:H_JU}
\end{equation}
where $\hat{n}_{1}^{\alpha}=\hat{a}_{\alpha}^{\dagger}\hat{a}_{\alpha}$,
and similarly for $k=2$. Here and in the following $\sum_{\alpha}$ is a shorthand notation for $\sum_{\alpha=L,R}$. We have used the following constants
\begin{eqnarray}
E_{k}&=&\int\!
d\mathbf{r}\,\psi_{k}^{\alpha *}(\mathbf{r})
\left(-\frac{\hbar^{2}}{2m}\nabla^{2}+U(\mathbf{r})\right)
\psi_k^\alpha(\mathbf{r}),\nonumber\\
\sigma&=&\frac{\Omega_R}{2}\int\nonumber\!
d\mathbf{r}\,\psi_{k}^{\alpha *}(\mathbf{r})\psi_k^\alpha(\mathbf{r}),\nonumber\\
J_{k}&=&-\int \!d\mathbf{r}\,\psi_{k}^{\alpha *}(\mathbf{r})
\left(-\frac{\hbar^{2}}{2m}\nabla^{2}+U(\mathbf{r})\right)
\psi_{k}^{\alpha'}(\mathbf{r}),\nonumber\\
U_{k}&=&g_{k}\int\! d\mathbf{r}|\psi_{k}^\alpha (\mathbf{r})|^{4}
\;,\end{eqnarray}
and $\varepsilon_k=E_k\pm\sigma $, where the minus sign
holds for $k=2$.
The inter-species term is
\begin{eqnarray}
\label{eq:Hintersepcies}
\hat{H}_{12} & = & \sum_{\alpha}
(S_{12}^{\alpha}+i \ D_{12}^\alpha)\hat{a}_{\alpha}^{\dagger}\hat{b}_{\alpha}
+\sum_{\alpha\ne\alpha'}\bar S_{12}^{\alpha,\alpha'}\hat{a}_{\alpha}^{\dagger}\hat{b}_{\alpha'}\nonumber\\
&+&\sum_{\alpha}(S_{21}^{\alpha}+i \ D_{21}^\alpha)\hat{b}_{\alpha}^{\dagger}\hat{a}_{\alpha}
 +  \sum_{\alpha\ne \alpha'}\bar S_{21}^{\alpha,\alpha'}\hat{b}_{\alpha}^{\dagger}\hat{a}_{\alpha'}\nonumber\\
& +& U_{12}\sum_\alpha \hat{n}_1^{\alpha}\hat{n}_2^{\alpha},
\end{eqnarray}
with
\begin{eqnarray}
U_{12}\!&=&\!g_{12}\int d\mathbf{r}\psi_{1}^{\alpha *}(\mathbf{r})\psi_{2}^{\alpha *}(\mathbf{r})\psi_2^\alpha(\mathbf{r})\psi_{1}^\alpha(\mathbf{r}),
\label{eq:AppCoeffs1}\\
S_{12}^\alpha\!&=&\!\int \!\!d\mathbf{r}\psi_{1}^{\alpha *}\hbar\left[ v_{R}\left(-\partial_{x}+i\partial_{y}\right) \!+ \! v_{D}\left(-\partial_{x}-i\partial_{y}\right)\right]\psi_{2}^\alpha,
\nonumber\\
S_{21}^\alpha\!&=&\!\int \!\!d\mathbf{r}\psi_{2}^{\alpha *}\hbar\left[ v_{R}\left(\partial_{x}+i\partial_{y}\right) \!+ \! v_{D}\left(\partial_{x}-i\partial_{y}\right)\right]\psi_{1}^\alpha,
\nonumber\\
\bar S_{12}^{\alpha,\alpha'}\!&=&\!\int \!\!d\mathbf{r}\psi_{k}^{\alpha *}\hbar\left[ v_{R}\left(-\partial_{x}+i\partial_{y}\right) \!+ \! v_{D}\left(-\partial_{x}-i\partial_{y}\right)\right]\psi_{k'}^{\alpha'},
\nonumber\\
\bar S_{21}^{\alpha,\alpha'}\!&=&\!\int \!\!d\mathbf{r}\psi_{2}^{\alpha *}\hbar\left[ v_{R}\left(\partial_{x}+i\partial_{y}\right) \!+ \! v_{D}\left(\partial_{x}-i\partial_{y}\right)\right]\psi_{1}^{\alpha'},
\nonumber\\
D_{12}^{\alpha}\!&=&\! -\int d\mathbf{r}\psi_{1}^{\alpha *}\frac{\delta}{2}\psi_{2}^{\alpha},\,\,\,\,D_{21}^{\alpha}= \int d\mathbf{r}\psi_{2}^{\alpha *}\frac{\delta}{2}\psi_{1}^{\alpha}.\nonumber
\end{eqnarray}
Non-zero inter-well interacting terms have been neglected, as commonly
assumed in standard two- or four-mode Hamiltonians~\cite{1999SpekkensPRA,2011GarciaMarchPRA,2012GarciaMarchFiP}.
This Hamiltonian conserves the number of atoms as it commutes with
the total number operator $\hat N=\sum_{k,\alpha} \hat n_k^\alpha$. For
$v_{R}=v_{D}=v$ the coefficients given in Eqs.~(\ref{eq:AppCoeffs1}) are the following
\begin{eqnarray}
S_{12}^{\alpha}\!&=&\!-2\hbar v\int d\mathbf{r}\, \psi_{1}^{\alpha *}(\mathbf{r})\partial_{x}\psi_{2}^\alpha(\mathbf{r}),
\\
S_{21}^{\alpha}\!&=&\!2\hbar v\int d\mathbf{r}\, \psi_{2}^{\alpha *}(\mathbf{r})\partial_{x}\psi_{1}^\alpha(\mathbf{r}),
\nonumber\\
\bar{S}_{12}^{\alpha,\alpha'}\!&=&\!-2\hbar v\int d\mathbf{r}\, \psi_{1}^{\alpha *}(\mathbf{r})\partial_{x}\psi_{2}^{\alpha'}(\mathbf{r})
\nonumber\\
\bar{S}_{21}^{\alpha,\alpha'}\!&=&\!2\hbar v\int d\mathbf{r}\, \psi_{2}^{\alpha *}(\mathbf{r})\partial_{x}\psi_{1}^{\alpha'}(\mathbf{r}),
\nonumber\\
D_{12}^{\alpha}\!&=&\! -\frac{\delta}{2}\! \int\! d\mathbf{r}\, \psi_{1}^{\alpha *}(\mathbf{r})\psi_{2}^\alpha(\mathbf{r}),\,
D_{21}^{\alpha}\! =\! \frac{\delta}{2}\! \int\! d\mathbf{r}\, \psi_{2}^{\alpha *}(\mathbf{r})\psi_{1}^\alpha(\mathbf{r}).
\nonumber
\end{eqnarray}
By integrating by parts, we get that
$S_{12}^{\alpha}=S_{21}^{\alpha}$ and $\bar{S}_{12}^{\alpha,\alpha'}=\bar{S}_{21}^{\alpha',\alpha}$. We also notice that $\bar{S}_{12}^{L,R}=-\bar{S}_{12}^{R,L}$ and that  $D_{12}^{\alpha}=-D_{21}^{\alpha}$. Then, by calling $S=S_{12}^{\alpha}$, $S_+=\bar{S}_{12}^{L,R}=\bar{S}_{21}^{R,L}$, $S_-=\bar{S}_{12}^{R,L}=\bar{S}_{21}^{L,R}$, and
$\tilde \delta=D_{12}^{\alpha}$, we can write Eq.~(\ref{eq:Hintersepcies}) as
\begin{eqnarray}
\label{eq:Hintersepcies1}
\hat{H}_{12} & = & \sum_{\alpha}
(S+i  \tilde \delta)\hat{a}_{\alpha}^{\dagger}\hat{b}_{\alpha}+\mbox{h.c.}+ U_{12}\sum_{\alpha} \hat{n}_1^{\alpha}\hat{n}_2^{\alpha}\nonumber\\
&+&S_+\hat{a}_{L}^{\dagger}\hat{b}_{R}+\mbox{h.c.}+S_-\hat{b}_{L}^{\dagger}\hat{a}_{R}+\mbox{h.c.}
\end{eqnarray}
We have checked that this Hamiltonian conserves the number of atoms. Notice that there are four different processes that interchange atoms between both species. The first two, associated to $S$ and $\tilde \delta$, interchange atoms between $k$ and $l$ components located in the same well. The third term, associated to $S_+$ interchanges atoms of $k$ component located in the left well and atoms of $l$ component located in the right well. The last term, $S_-$, transforms atoms of $k$ in the right well to atoms of $l$ in the left well, and vice-versa.
From the Heisenberg equations of motion for the operators $\hat a_{\alpha}$ and $\hat b_{\alpha}$
\begin{equation}
i\hbar\frac{d \hat a_{\alpha}}{dt}=\left[\hat a_{\alpha},\hat H\right]\,\,\,\mbox{and}\,\, i\hbar\frac{d\hat b_{\alpha }}{dt}=\left[\hat b_{\alpha},\hat H\right]
\end{equation}
one can obtain
\begin{eqnarray}
\label{eq:appeqoperatora}
i\hbar\frac{d\hat a_{\alpha}}{dt} & = & -J_{1}\hat a_{\alpha'}+U_{1}\hat n_{1}^\alpha \hat a_{\alpha}+\varepsilon_{1}\hat a_{\alpha}+U_{12}\hat n_{2}^\alpha \hat a_{\alpha}\nonumber\\
&-&S\hat b_{\alpha}-\bar{S}_\pm\hat b_{\alpha'}-i\tilde \delta\hat b_{\alpha},\\
\label{eq:appeqoperatorb}
i\hbar\frac{d\hat b_{\alpha}}{dt} & = & -J_{2}\hat b_{\alpha'}+U_{2}\hat n_{2}^\alpha\hat b_{\alpha}+\varepsilon_{2}\hat b_{\alpha}+U_{12}\hat n_{1}^\alpha\hat b_{\alpha}\nonumber\\
&-&S\hat a_{\alpha}-\bar{S}_\mp\hat a_{\alpha'}+i\tilde \delta\hat a_{\alpha},
\end{eqnarray}
and their corresponding Hermitian conjugates. The upper(lower) sign in the $S_\pm$ coefficients apply to $\alpha=L(R)$. To reconcile the definition of
the coefficients with the one given in Eqs.~(\ref{parameters})
we have redefined all coefficients as positive, and therefore we have written the minus sign inside their definitions explicitly in the equations.
We assume that the creation/annihilation operators behave
as $c$-numbers, that is $a_{\alpha}=\sqrt{N_{1}^\alpha}e^{i\phi_{1}^\alpha}$ where
$N_{1}^\alpha=|a_\alpha|^{2}$ is the number of particles of species 1 and
$\phi_{1}^\alpha$ is a phase
(similarly $b_\alpha=\sqrt{N_{2}^\alpha}e^{i\phi_{2}^\alpha}$). After some
algebra, the equations of motion for the number of particles,
Eq.~(\ref{eq:Nkalpha1}), and phases, Eq.~(\ref{eq:thekalpha1}), are
obtained from the equations of
motion~(\ref{eq:appeqoperatora})-(\ref{eq:appeqoperatorb}). Since we assumed that $m_1=m_2=m$, and therefore one can use the same localized function for the two components, we obtain that $S=0$ and $\epsilon_1=\epsilon_2$, and consequently the corresponding terms are absent in Eqs.~(\ref{eq:Nkalpha1})-(\ref{eq:thekalpha1}).

\end{document}